\documentclass[12pt]{article}

\usepackage{graphicx}
\usepackage{moreverb}
\usepackage{animate}
\usepackage{multimedia}
\usepackage{graphicx,psfrag,epsf}
\usepackage{enumerate}
\usepackage{ulem}
\usepackage{xcolor}
\usepackage{wrapfig}
\usepackage{color}
\usepackage{amssymb}
\usepackage{caption}
\usepackage[caption=false]{subfig}
\usepackage{authblk}
\usepackage{mathtools}
\usepackage{cite}
\usepackage{natbib}
\usepackage[export]{adjustbox}

\definecolor{dgreen}{HTML}{007000}

\DeclarePairedDelimiter{\ceil}{\lceil}{\rceil}

\newcommand{\bs}[1]{\boldsymbol{#1}}

\begin{document}

\title{Local projections for high-dimensional outlier detection}


\author{Thomas Ortner$^1$\footnote{thomas.ortner@tuwien.ac.at}, Peter Filzmoser$^1$, Maia Zaharieva$^1$, Sarka Brodinova$^2$ and Christian Breiteneder$^2$\\
TU Wien\\
$^1$Institute of Statistics and Mathematical Methods in Economics \\
$^2$Institute of Software Technology and Interactive Systems
}

%\authorrunning{Ortner et.al.} % if too long for running head

%\institute{T. Ortner \and P. Filzmoser \and S. Brodinova  \at
%Institute of Statistics and Mathematical Methods in Economics, 
%Vienna University of Technology \\
%              %\email{thomas.ortner@tuwien.ac.at}
%           M. Zaharieva \and C. Breiteneder \at
%Institute of Software Technology and Interactive Systems,  Vienna University of Technology
%}


\maketitle

 \begin{abstract}
In this paper, we propose a novel approach for outlier detection, called local projections, which is based on concepts of Local Outlier Factor (LOF)~\citep{breunig2000lof} and RobPCA~\citep{hubert2005robpca}. By using aspects of both methods, our algorithm is robust towards noise variables and is capable of performing outlier detection in multi-group situations. We are further not reliant on a specific underlying data distribution. 

For each observation of a dataset, we identify a local group of dense nearby observations, which we call a core, based on a modification of the k-nearest neighbours algorithm. By projecting the dataset onto the space spanned by those observations, two aspects are revealed. First, we can analyze the distance from an observation to the center of the core within the projection space in order to provide a measure of quality of description of the observation by the projection. Second, we consider the distance of the observation to the projection space in order to assess the suitability of the core for describing the outlyingness of the observation. These novel interpretations lead to a univariate measure of outlyingness based on aggregations over all local projections, which outperforms LOF and RobPCA as well as other popular methods like PCOut~\citep{filzmoser2008outlier} and subspace-based outlier detection~\citep{kriegel2009outlier} in our simulation setups. Experiments in the context of real-word applications employing datasets of various dimensionality demonstrate the advantages of local projections.
\end{abstract}

\section{Introduction}
\sloppy 

Classical outlier detection approaches in the field of statistics are experiencing multiple problems in the course of the latest developments in data analysis. The increasing number of variables, especially non-informative noise variables, combined with complex multivariate variable distributions makes it difficult to compute classical critical values for flagging outliers. This is mainly due to singular covariance matrices, distorted distribution functions and therefore skewed critical values \citep[e.g.][]{aggarwal2001outlier}. At the same time, outlier detection methods from the field of computer science, which do not necessarily rely on classical assumptions such as normal distribution, enjoy an increase in popularity even though their application is commonly limited due to large numbers of variables or flat data structures (more variables than observations). These observations motivated the proposed approach for outlier detection incorporating aspects from two popular methods: the Local Outlier Factor (LOF)~\citep{breunig2000lof}, originating in the computer science, and RobPCA, a robust principal component analysis-based (PCA) approach for outlier detection coming from the field of robust statistics~\citep{hubert2005robpca}. The core aim of the proposed approach is to measure the outlyingness of observations avoiding any assumptions on the underlying data distribution and being able to cope with high-dimensional datasets with fewer observations than variables (flat data structures).

LOF avoids any assumptions on the data distribution by incorporating a k-nearest neighbour algorithm. Within groups of neighbours, it evaluates whether or not an observation is located in a similar density as its neighbours. Therefore, multi-group structures, skewed distributions, and other obstacles have minor impact on the method as long as there are enough observations for modelling the local behaviour. On the contrary, RobPCA uses a robust approach for modelling the majority of observations, which are assumed to be normally distributed. It uses a projection on a subspace based on this majority. In contrast to most other approaches, RobPCA does not only investigate this subspace but also the orthogonal complement, which reduces the risk of missing outliers due to the projection procedure. 

The proposed approach aims at combining these two aspects by defining projections based on the local neighbourhood of an observation where no reliable assumption about the data structure can be made and by considering the concept of the orthogonal complement similar to RobPCA. The approach of local projections is an extension of \textit{Guided projections for analyzing the structure of high-dimensional data}~\citep{ortner2017guided}. We identify a subset of observations locally, describing the structure of a dataset in order to evaluate the outlyingness of other nearby observations. While guided projections create a sequence of projections by exchanging one observation by another and re-project the data onto the new selection of observations, in this work, we re-initiate the subset selection in order to cover the full data structure as good as possible with $n$ local descriptions, where $n$ represents the total number of observations. We discuss how outlyingness can be interpreted with regard to local projections, why the local projections are suitable for describing the outlyingness of an observation, and how to combine those projections in order to receive an overall outlyingness estimation for each observation of a dataset. 

The procedure of utilizing projections linked to specific locations in the data space has the crucial advantage of avoiding any assumptions about the distribution of the analyzed data as utilized by other knn-based outlier detection methods as well \citep[e.g.][]{kriegel2009outlier}. Furthermore, multi-group structures do not pose a problem due to the local investigation. 
%This is possible due to our novel interpretation of outlyingness which, to the best of our knowledge, has not been considered before.

We compare our approach to related and well-established methods for measuring outlyingness. Besides RobPCA and LOF, we consider PCOut~\citep{filzmoser2008outlier}, an outlier detection method focusing on high-dimensional data from the statistics, KNN~\citep{campos2016evaluation}, since our algorithm incorporates knn-selection similar to LOF, subspace-based outlier detection (SOD)~\citep{kriegel2009outlier}, a popular subspace selection method from the computer science and Outlier Detection in Arbitrary Subspaces (COP)~\citep{kriegel2012outlier}, which follows a similar approach but has difficulties when dealing with flat  data structures. Our main focus in this comparison is exploring the robustness towards an increasing number of noise variables.

The paper is structured as follows: Section~\ref{sec:locproj} provides the background for a single local projection including a demonstration example. We then provide an interpretation of outlyingness with respect to a single local projection and a solution for aggregating the information based on series of local projections in Section~\ref{sec:outlyingness}. Section~\ref{sec:evaluation} describes all methods used in the comparison, which are then applied in two simulated settings in Section~\ref{sec:simulations}. Finally, in Section~\ref{sec:application}, we show the impact on three real-world data problems of varying dimensionality and group structure before we provide a brief discussion on the computation time in Section~\ref{sec:computational}. 
We conclude with a discussion in Section \ref{sec:conclusion}.

\section{Local projections}
\label{sec:locproj}

Let $\bs X$ denote a data matrix with $n$ rows (observations) and $p$ columns (variables) drawn from a $p$-dimensional random variable $X$, following a non-specified distribution function $F_X$. We explicitly consider the possibility of $p>n$ to emphasize the situation of high-dimensional low sample size data referred to as flat data, which commonly emerges in modern data analysis problems. We assume that $F_X$ represents a mixture of multiple distributions $F_{X_1}, \dots, F_{X_q}$, where the number of sub-distributions $q$ is unknown.  The distributions are unspecified and can differ from each other. However, we assume that the distributions are continuous. Therefore, no ties are present in the data, which is a reasonable assumption especially for a high number of variables. In the case of ties, a preprocessing step, excluding ties can be applied in order to meet this assumption. An outlier in this context is any observation, which deviates from each of the groups of observations associated with the $q$ sub-distributions.

Our approach for evaluating the outlyingness of observations is based on the concept of using robust approximations of $F_X$, which do not necessarily need to provide a good overall estimation of $F_X$ on the whole support but only of the local neighborhood of each observation. Therefore, we aim at estimating the local distribution around each observation $\bs x_i$, for $i = 1, \dots ,n$, not by all available observations but by a small subset, which {is located close to} $\bs x_i$. 

We limit the number of observations included in the local description in order to avoid the influence of inhomogenity in the distribution (e.g. multimodal distributions or outliers being present in the local neighbourhood) of the underlying random variable.

For complex problems, especially high-dimensional problems, such approximations are difficult to find. We use projections onto groups of observations locally describing the distribution. Therefore, we start by introducing the concept of a local projection, which will then be used as one such approximation before describing a possibility of combining those local approximations. In order to provide a more practical incentive, we demonstrate the technical idea in a simulated example throughout the section.

\subsection{Definition of local projections}
\label{sec:LocOut}

Let $\bs y$ denote one particular observation of the data matrix $\bs X = (\bs x_1, \dots , \bs x_n)'$, where $\bs x_i=(x_{i1} \dots x_{ip})'$. For any such $\bs y$, we can identify its $k$ nearest neighbours using the Euclidean distance between $\bs y$ and $\bs x_i$, denoted by $d(\bs y, \bs x_i)$ for all $i=1, \dots ,n$:
\begin{equation}
knn(\bs y) = \{ \bs x_i: d(\bs y, \bs x_i) \leq d_k \}, \label{eq:knn}
\end{equation}
where $d_k$ is the $k$-smallest distance from $\bs y$ to any other observation in the dataset. 

Using the strategy of robust estimation, we consider {a subset of} $\ceil{\alpha \cdot k}$ observations from $knn(\bs y)$ for the description of the local distribution, where $\alpha$ represents a trimming parameter describing the proportion of observations, which are assumed to be non-outlying in any $knn$. Here, $\ceil{c}$ denotes the {smallest} integer $\geq c$. The parameter $\alpha$ is usually set to $0.5$ in order to avoid neighbors that are heterogeneous (e.g. due to outliers) but it can be adjusted if additional information about the specific dataset is available. By doing so, we reduce the influence of outlying observations, which would distort our estimation. The idea is to get the most dense group of $\ceil{\alpha \cdot k}$ observations, which we call the \textit{core} of the projection, initiated by $\bs y$, not including $\bs y$ itself. The center of this core is defined by
\begin{equation}
\bs x_0 = arg \min_{\bs x_i \in knn(\bs y) }\{ d_{(\ceil{\alpha \cdot k})}(\bs x_i) \} ,
\end{equation}
where  {$d_{(\ceil{\alpha \cdot k})}(\bs x_i)$} represents the $\ceil{\alpha \cdot k}$-largest distance between $\bs x_i$ and any other observation from $knn(\bs y)$. The observation $\bs x_0$ can be used to define the $core$ of a local projection initiated by $\bs y$:
\begin{align}
core(\bs y) = \{ \bs x_i: & d(\bs x_0, \bs x_i) < d_{(\ceil{\alpha \cdot k})}(\bs x_0)  \wedge  \nonumber \\ 
 & \bs x_i \in knn(\bs y) {\wedge \bs x_i \neq \bs y} \}
\end{align}

In order to provide an intuitive access to the proposed approach,
we explain the concept of local projections for a set of simulated observations. In this example, we use 200 observations drawn from a two-dimensional normal distribution. The original observations and the procedure of selecting the $core(\bs y)$ are visualized in Figure 1: The red observation was manually selected to initiate our local projection process and refers to $\bs y$.  It can be exchanged by any other observation. However, in order to emphasize the necessity of the second step of our procedure, we selected an observation off the center. The blue observations are the $k=20$ nearest neighbours of $\bs y$ and the filled {blue circles represent} the core of $\bs y$ using $\alpha=0.5$. We note that the observations of $core(\bs y)$ tend to be closer to the center of the distribution than $\bs y$ itself, since  we can expect an increasing density towards the center of the distribution, which likely leads to more dense groups {of} observations.

\begin{figure}[tbh]
\centering
\includegraphics[width=0.475\linewidth]{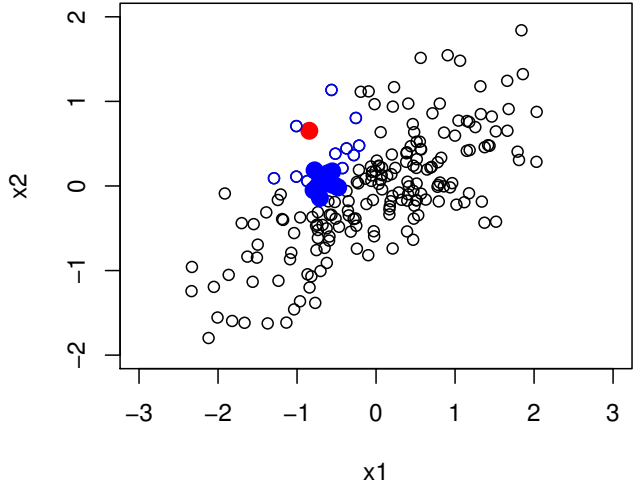}
\caption{Visualization of the $core$-selection process. The red observation represents the initiating observation $\bs y$. The blue observations represent $knn(\bs y)$ and the filled blue observations represent $core(\bs y)$. $x_0$ itself is not visualized but it is known to be an element of $core(\bs y)$.}
\label{fig:selection}
\end{figure}

A projection onto the space, spanned by the observations contained in $core(\bs y)$, provides a description of the similarity between any observation and the core, which is especially of interest for $\bs y$ itself. Such a projection can efficiently be computed using the singular value decomposition (SVD) of the matrix of observations in $core(\bs y)$, centered and scaled with respect to the core itself. In order to estimate the location and scale parameters for scaling the data, we can apply classical estimators on the core preserving robustness properties, since the observations have been included into the core in a robust way.
\begin{align}
\bs X_{core(\bs y )}  =&  (\bs x_{\bs y,1}, \dots, \bs x_{\bs y,{\ceil{\alpha \cdot k}}})' \\ 
\nonumber
&\bs x_{\bs y,j} \in core(\bs y) \hspace{25pt} \forall j \in \{1, \dots, \ceil{\alpha \cdot k} \} \\ 
\hat{\bs \mu}_{\bs y}  = & \frac{1}{\ceil{\alpha \cdot k}} \sum_{\bs x_i \in core(\bs y)} \bs x_i  \label{eq:center} \\
\bs{\hat{ \sigma}}_{\bs y} =& \left( \sqrt{Var(x_{\bs y,11}, \dots, x_{\bs y,\ceil{\alpha \cdot k}1} )}, \dots,  \right. \nonumber \\
& \left.  \hspace{30pt} \sqrt{Var(x_{\bs y,1p}, \dots, x_{\bs y,\ceil{\alpha \cdot k}p} )}\right)' \\ 
\nonumber
= & (\hat{\sigma}_{\bs y,1}, \dots, \hat{\sigma}_{\bs y,p})' ,
\end{align}
{where $Var$ denotes the sample variance.} Using $\hat{ \bs \mu }_{\bs y}$, the centered observations are given by
\begin{equation}
\bs x^c_{\bs y} =  (x^c_{\bs y, 1}, \dots x^c_{\bs y, p} )' = \bs x_{\bs y} - \bs{\hat{\mu}_{\bs y }}, \\
\label{eq:center}
\end{equation}
which can be used to provide the centered and column-wise scaled data matrix with respect to the core of $\bs y$:
 \begin{align}
\tilde{\bs X}_{\bs y} = &\left( 
\left( \frac{x_{\bs y,11}^c}{\hat \sigma_{\bs y,1}}, \dots, \frac{x_{\bs y,1p}^c}{\hat \sigma_{\bs y, p}} \right)',
\dots,\right.  \nonumber \\
& \hspace{10pt} \left. \left( \frac{x_{\bs y \ceil{\alpha \cdot k}1}^c}{\hat \sigma_{\bs y,1}}, \dots, \frac{x_{\bs y,\ceil{\alpha \cdot k}p}^c}{\hat \sigma_{\bs y ,p}} \right)'
 \right)'
 \end{align}
Based on $\tilde{ \bs  X}_{\bs y}$, we provide a projection onto the space spanned by the observations of $core(\bs y)$ by $\bs V_{\bs y}$ from the SVD of $\tilde{\bs  X}_{\bs y}$,
 \begin{equation}
 \tilde{\bs  X}_{\bs y} =  \bs U_{\bs y} \bs D_{\bs y}  \bs V_{\bs y}' \label{eq:svd} .
 \end{equation}
  
Any observation $\bs x$ can be projected onto the projection space by centering with $\bs{\hat{\mu}_{\bs y}}$, scaling with $\bs{\hat{ \sigma}}_{\bs y}$, and applying the linear transformation $\bs V'_{\bs y}$. The projection of the whole dataset is given by $\tilde{ \bs X }_{\bs y} \bs V_{\bs y}$. We refer to the projected observations as the representation of observations in the \textit{core space} of $\bs y$. Since the dimension of the core space is limited by $\ceil{\alpha \cdot k}$, in any case where $p>\ceil{\alpha \cdot k}$ holds and $\bs X_{core(\bs y)}$ is of full rank, a non-empty orthogonal complement of this core space exists. Therefore, any observation $\bs x$ consists of two representations, the core representation $\bs x_{core(\bs y)}$ given the core space,
\begin{equation}
\bs x_{core(\bs y)}  =  \bs V'_{\bs y} \left( \frac{x_1^c}{\hat \sigma_{\bs y,1}}, \dots, \frac{x_p^c}{\hat \sigma_{\bs y, p}} \right)' ,
\end{equation}
where $\bs x^c = (x_1^c, \dots , x_p^c)'$ is computed as defined in Equation~\eqref{eq:center} and the orthogonal representation $\bs x_{orth(\bs y)}$ given the orthogonal complement of the core space,
\begin{equation}
\bs x_{orth(\bs y)} = \bs x^c -  \bs V_{\bs y}\bs x_{core(\bs y)} .
\label{eq:xorth}
\end{equation}

\begin{figure}[tbh]
\subfloat[]{
    \label{fig:selection2a}
    \includegraphics[width=0.475\linewidth]{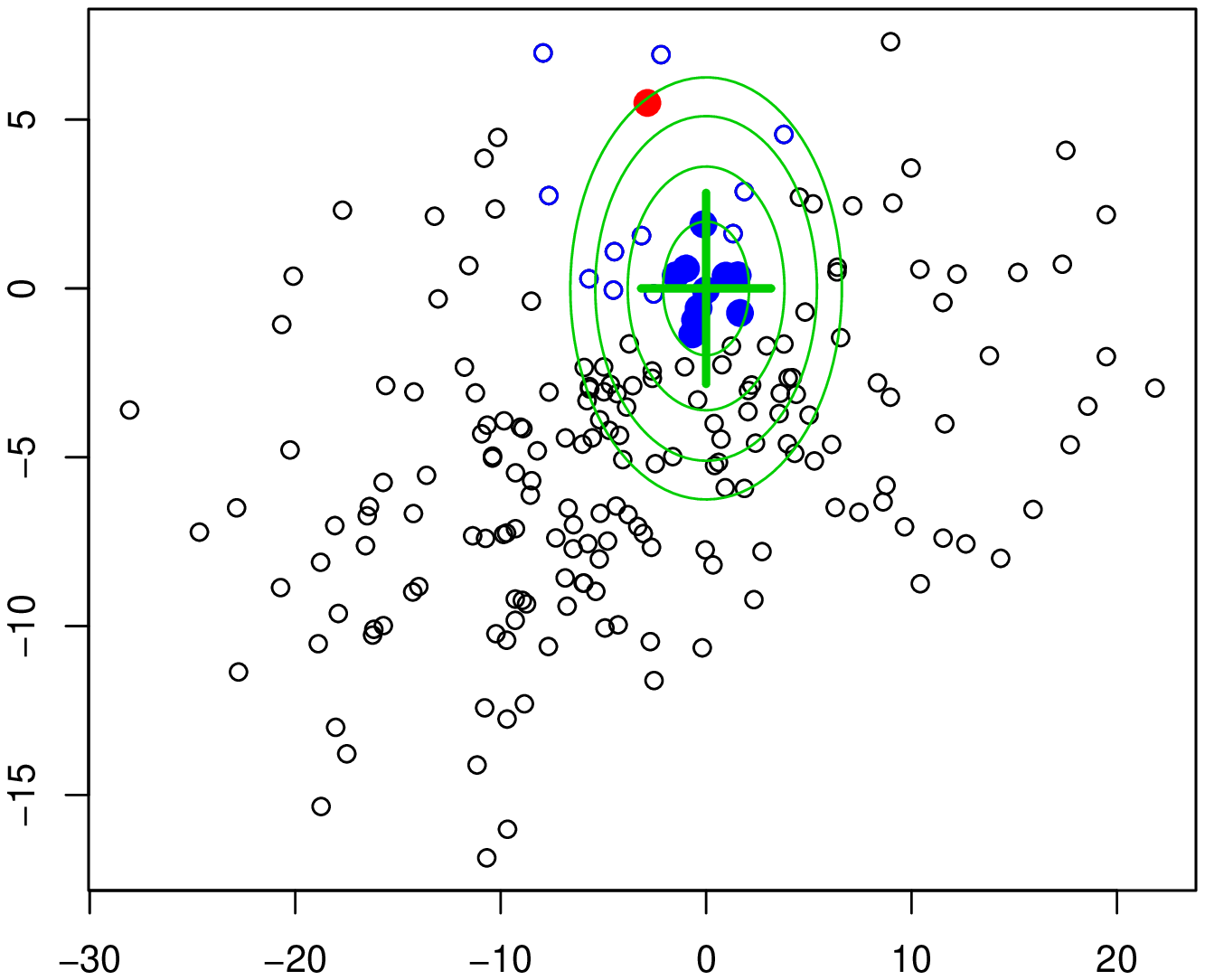} }
\subfloat[]{
    \label{fig:selection2b}
    \includegraphics[width=0.475\linewidth]{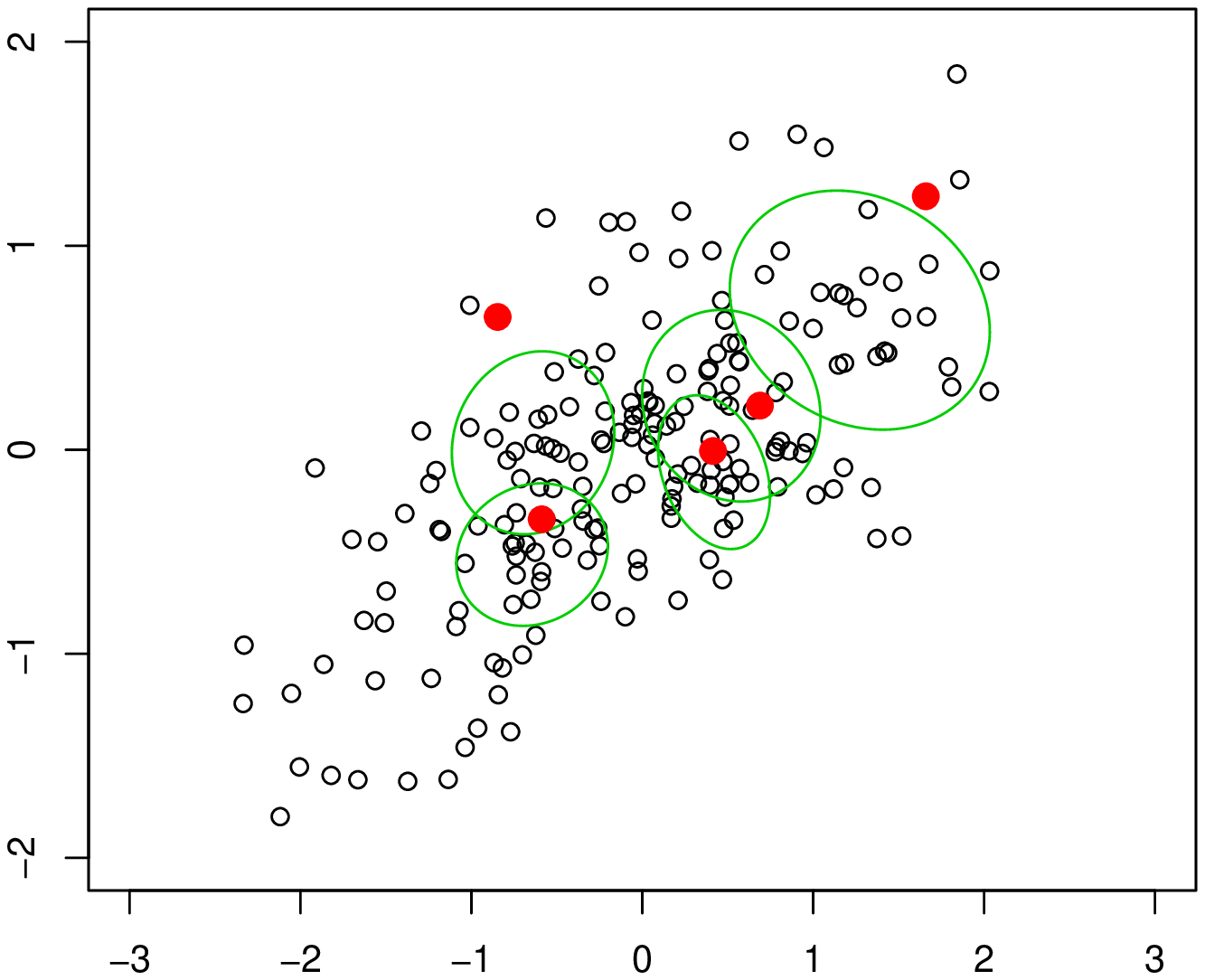} }
    \caption{Plot (a) provides a visualization of the transformed observations from Figure~\ref{fig:selection}. The red observation represents the initiating observation $\bs y$. The blue observations represent $knn(\bs y)$ and the filled blue observations represent $core(\bs y)$. The green ellipses represent the covariance structure estimated by the core observations representing the local distribution. Plot (b) uses the same representation as Figure~\ref{fig:selection} but shows the concept of multiple local projections initiated by different observations marked as red dots. Each of the core distances represented by green ellipses refers to the same constant value taking the different covariance structures of the different cores into account.}
    \label{fig:selection2}
  \end{figure}

Figure~\ref{fig:selection2a} shows the representation of our 200 simulated observations in the core space. Note that in this special case, the orthogonal representation is constantly $\bs 0$ due to the non-flat data structure of the core observations ($p<k$). We further see that the center of the core is now located in the center of the coordinate system.

Given a large enough number of observations and a small enough dimension of the sample space, we can approximate  $F_X$ with arbitrary accuracy given any desired neighborhood. However, in practice, the quality of this approximation is limited by a finite number of observations. Therefore, it depends on various aspects like the size of $d_k$ and $d_{\ceil{\alpha \cdot k}}$ and, thus, the approximation is always limited by the restrictions imposed by the properties of the dataset. Especially the behavior of the core observations will, in practice, significantly deviate from the expected distribution with increasing $d_{\ceil{\alpha \cdot k}}$.

In order to take this local distribution into account, it is useful to include the properties of the core observations in the core space into the distance definition within the core space. A more advantageous way to measure the deviation of core distances from the center of the core than using Euclidean distances is the usage of Mahalanobis distances \citep[e.g.][]{de2000mahalanobis}.
For the projection space, an orthogonal basis is defined by the left eigenvectors of the SVD from Equation~\eqref{eq:svd}, while the singular values given by the diagonal of the matrix $\bs D_{\bs y}$ provide the standard deviation for each direction of the projection basis. Therefore, weighting the directions of the Euclidean distances with the inverse singular values directly leads to Mahalanobis distances  in the core space, which take the variation of each direction into account:

\begin{equation}
CD_{\bs y} (\bs x) = \sqrt{ \frac{\bs x'_{core(\bs y)} \bs D_{\bs y}^{-1} 
 \bs x_{core(\bs y)}}{min(\ceil{\alpha \cdot k}-1, p )  }  }
\label{eq:cd} 
\end{equation}.

The computation of core distances can be derived from Figure~\ref{fig:selection2a}. The green cross in the center of the coordinate system refers to the (projected) left singular vectors of the SVD. We note that the scale of the two axes in Figure~\ref{fig:selection2} differ appreciably. The green ellipses represent Mahalanobis distances based on the variation of the two orthogonal axes, which provide a more suitable measure for describing the distribution locally.

The distances of the representation in the orthogonal complement of the core cannot be rescaled as in the core space. All observations from the core, which are used to estimate the local structure, i.e. to span the core space, are fully located in the core space. Therefore, their orthogonal complement is equal to $\bs 0$:
\begin{equation}
\bs x_{orth(\bs y)} = \bs 0 \hspace{15pt} \forall \bs x \in core(\bs y)
\end{equation}

Since no variation in the orthogonal complement is available, we cannot estimate the rescaling parameters for the orthogonal components. Therefore, we directly use the Euclidean distances in order to describe the distance from any observation $\bs x$ to the core space of $\bs y$. We will refer to this distance as orthogonal distance ($OD$).
\begin{equation}
OD_{\bs y} (\bs x) = || \bs x_{orth(\bs y)} ||
\label{eq:od} 
\end{equation}

The two measures for similarity, $CD$ and $OD$, are inspired by the score and the orthogonal distance of~\cite{hubert2005robpca}. In contrast to~\cite{hubert2005robpca}, we do not try to elaborate critical values for $CD$ and $OD$ to directly decide if an observation is an outlier. Such critical values always depend on an underlying normal distribution and on the variation of the core and the orthogonal distances of the core observations. Instead, we aim at providing multiple local projections in order to be able to estimate the degree of outlyingness for observations in any location of a data set. A core and its core distances can be defined for every observation. Therefore, a total of $n$ projections with core and orthogonal distances are available for analyzing the data structure. Figure~\ref{fig:selection2b} visualizes a small number (5) of such projections in order to demonstrate how the concept works in practice. The red observations are used as the initiating observations, the green ellipses represent core distances based on each of the 5 cores. Each core distance refers to the same constant value considering the different covariance estimations of each core. We see that observations closer to the boundary of the data
%which can also visually be identified as outliers, 
are described less adequately by their respective core, while other observations, close to the center of the distribution, are well described by multiple cores.

\section{Interpretation and utilization of local projections}
\label{sec:outlyingness}

Most subspace-based outlier detection methods, including PCA-based methods such as \textit{PCOut} \citep{filzmoser2008outlier} and projection pursuit methods \citep[e.g.][]{henrion2013casos}, focus on the outlyingness of observations within a single subspace only. The risk of missing outliers due to the subspace selection by the applied method is evident as the critical information might remain outside the projection space. RobPCA \citep{hubert2005robpca} is one of the few methods considering the distance to the projection space in order to monitor this risk as well.

We would like to use both aspects, distances within the projection space and to the projection space, to evaluate the outlyingness of observations as follows: The projection space itself is often used as a model, employed to measure the outlyingness of an observation. Since we are using a local knn-based description, we can not directly apply this concept as our projections are bound to a specific location defined by the cores. The core distance from the location of our projection rather describes whether an observation is close to the selected core. If this is the case, we can assume that the model of description (the projection represented by the projection space) fits the observation well. Therefore, if the observation is well-described, there should be little information remaining in the orthogonal complement leading to small orthogonal distances.

\begin{figure}[!hbt]
\centering
\subfloat[]{  \includegraphics[width=0.4\linewidth]{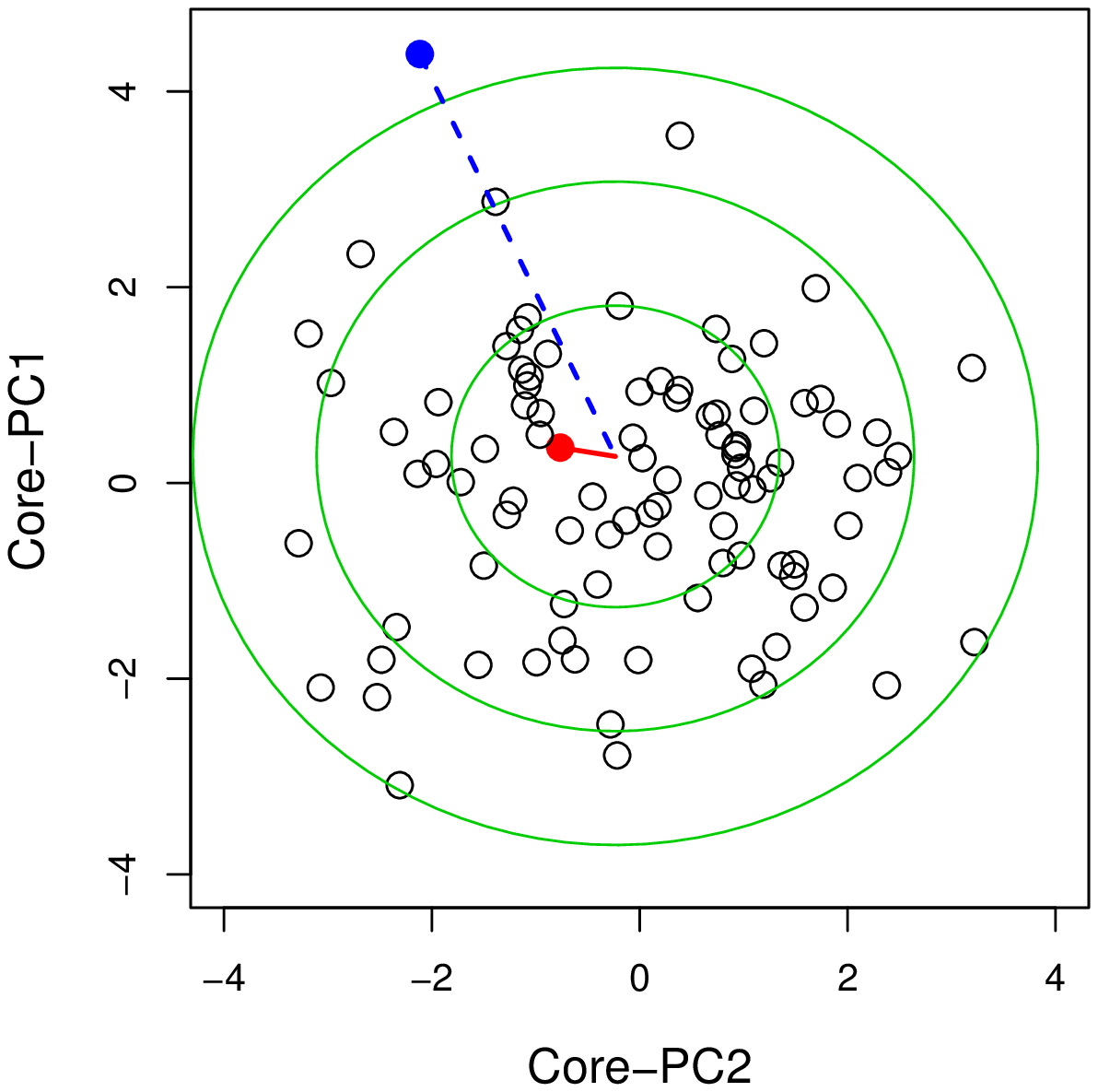} }
\subfloat[]{  \includegraphics[width=0.4\linewidth]{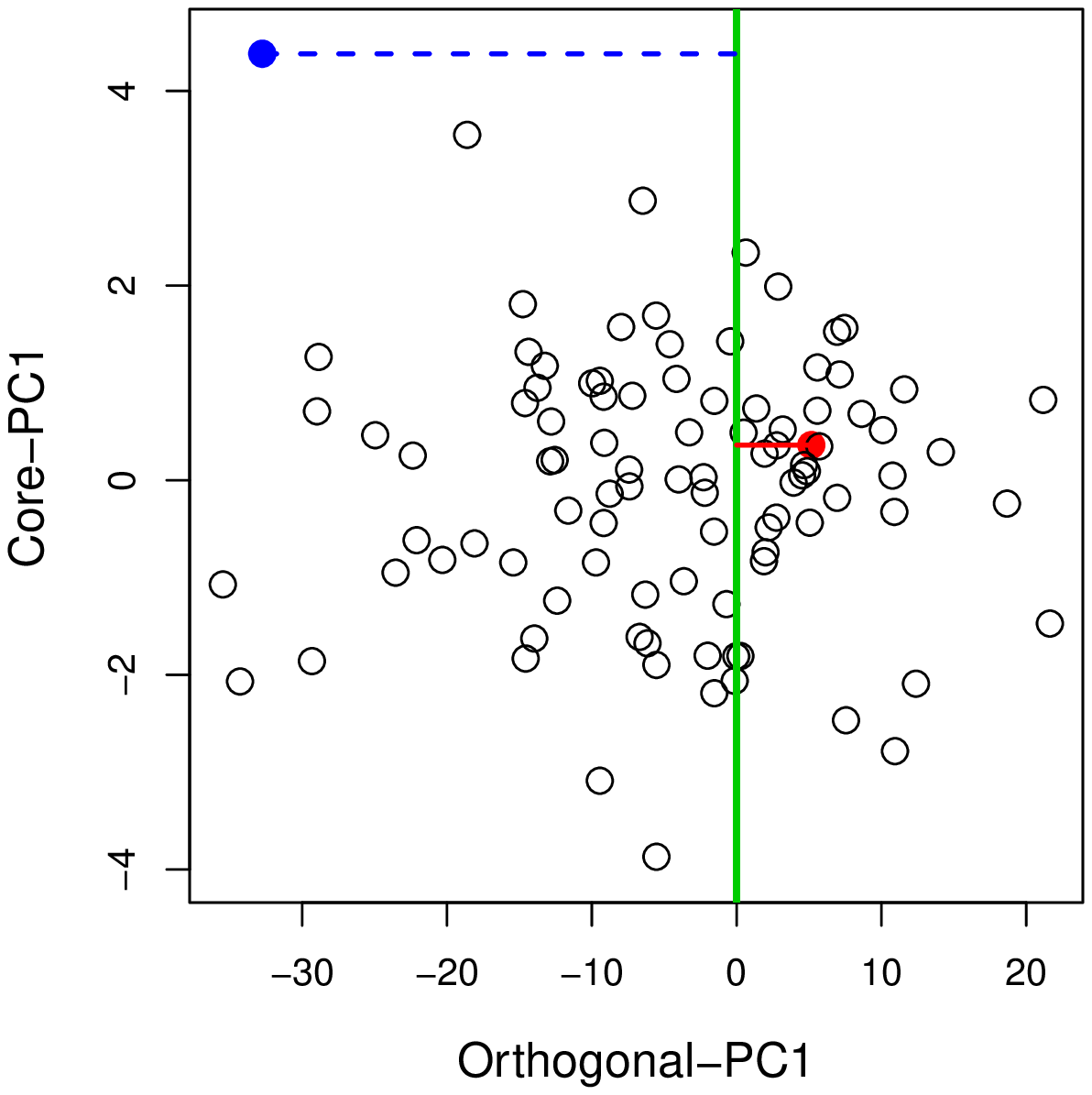} }
\caption{Visualization of orthogonal and core distances for a local projection of a multivariate 100-dimensional normal distribution. Plot (a) describes the core space by its first two principal components. The measurement of the core distances is represented by the green ellipses. Plot (b) includes the orthogonal distance. The vertical green line represents the projection space.}
\label{fig:distances1}
\end{figure}

We visualize this approach in Figure~\ref{fig:distances1} in two plots. Plot (a) shows the first two principal components of the core space and plot (b) the first principal component of the core and the orthogonal space respectively. In order to retrace our concept of interpreting core distances as the quality of the local description model and the core distances as a measure of outlyingness with respect to this description, we look at the two observations marked in red and blue. While the red observation is close to the center of our core as seen in plot (a), the blue one is located far off. Therefore, the blue observation is not as well described by the core as the red observation, which becomes evident when looking at the first principal component of the orthogonal complement in plot (b), where the blue observation is located far off the green line representing the projection space.

Note that this interpretation does not hold for core observations. This is due to the fact that the full information of core distances is located in the core space. With increasing $p$, the distance of all observations from the same group converges to a constant as shown in e.g. \cite{filzmoser2008outlier} for multivariate normal distributions. While this distance is completely represented in the core space for core observations, a proportion of distances from non-core observations will be represented in the orthogonal complement of the core space. Therefore, the probability of the core distance of a core observation being larger than the core distance of any other observation from the same group converges to $1$ with increasing $p$:

\begin{align}
\lim_{p \rightarrow \infty}P( CD_{\bs y}(\bs x_i) > CD_{\bs y}(\bs z) ) = 1,  \hspace{10pt}&  \bs x_i \in core(\bs y),  \label{eq:prob1}  \\ \nonumber 
& \bs z \notin core(\bs y) 
\end{align}

So far we considered a single projection, where we deal with a total of $n$ projections. Let $\mathcal X$ denote a set of $n$ observations $\{ \bs x_1,\dots, \bs x_n  \}$. Therefore, for each initializing observation $\bs x \in \mathcal X$, the core distance $CD_{\bs x}$ and the orthogonal distance $OD_{\bs x}$ are well-defined for all observations from $\mathcal X$. As motivated above, we want to measure the quality of local description using the core distances and the local outlyingness using the orthogonal distances. The smaller the core distance of an observation for a specific projection is, the more relevant this projection is for the overall evaluation of the outlyingness of this observation.  Therefore, we downweight the orthogonal distance based on the inverse core distances. In order to make the final outlyingness score comparable, we scale these weights by setting the sum of weights to $1$ for each local projection initiating observation $\bs y$:
\begin{equation}
 w_{\bs y}(\bs x) =  \left\{\begin{array}{lr}
 0,  & \bs x \in core(\bs y) \\
 \frac{ \frac{1}{ CD_{\bs y}(\bs x)}
  - \min\limits_{ \tilde{\bs z} \in \mathcal X}   \left(
 \frac{1}{CD_{\tilde{\bs z}}(\bs x)}  \right) 
  }{  \sum\limits_{\bs z \in \mathcal X} \left( \frac{1}{CD_{\bs z}(\bs x)} - \min\limits_{ \tilde{\bs z} \in \mathcal X}   \left(
 \frac{1}{CD_{\tilde{\bs z}}(\bs x)}  \right)   \right) }, & else
\end{array}\right.
\end{equation}

The scaled weights $w_{\bs y}$ make sure, that the sum of contributions by all available projections remains the same. Therefore, the sum of weighted orthogonal distances, corresponding to local outlyingness through local projections (LocOut),
 \begin{equation}
 LocOut(\bs x) = \sum\limits_{\bs y \in \mathcal X} \left( w_{\bs y}(\bs x ) \cdot OD_{\bs y} (\bs x) \right),
 \end{equation}
provides a useful, comparable measure of outlyingness for each observation. 

Note that this concept of outlyingness is limited to high-dimensional spaces. Whenever we analyze a space where $p\leq \ceil{\alpha \cdot k}$ holds, the full information of all observations will be located in the core space of each local projection. Therefore, for varying core distances, the orthogonal distance will always remain zero. Thus, the weighted sum of orthogonal distances can not provide any information on outlyingness unless there is information available in the orthogonal representation of observations.
 
\section{{Evaluation setup}}
\label{sec:evaluation}

In order to evaluate the performance of our proposed methodology, we compare it with related algorithms, namely LOF \citep{breunig2000lof},  Rob\-PCA \citep{hubert2005robpca}, PCOut \citep{filzmoser2008outlier},  {COP \citep{kriegel2012outlier}, KNN \citep{ramaswamy2000efficient}}, and SOD \citep{kriegel2009outlier}. Each of those algorithms tries to identify outliers in the presence of noise variables.  Some methods use a configuration parameter describing the dimensionality of the resulting subspace or the number of neighbours in a knn-based algorithm. In our algorithm, we use $\ceil{\alpha \cdot k}$ observations to create a subspace, which we employ for assessing the outlyingness of observations. In order to provide a fair comparison, the configuration parameters of each method are adjusted individually for each dataset: We systematically test different configuration values and report the best achieved performance for each method. Instead of outlier classification, we rather use each of the computed measures of outlyingness since not all methods provide cutoff values.
The performance itself is reported in terms of the commonly used area under the ROC Curve (AUC)~\citep{fawcett2006introduction}.

\subsection{Compared methods}

\textbf{Local Outlier Factor (LOF)}~\citep{breunig2000lof} is one of the main inspirations for our approach. The similarity of observations is described using ratios of Euclidean distances to k-nearest observations. Whenever this ratio is close to 1, there is a consistent group of observations and, therefore, no outliers. As for most outlier detection methods, no explicit critical value can be provided for LOF  \citep[e.g.][]{zimek2012survey, campos2016evaluation}. {In order to optimize the performance of LOF, we estimate the number of k-nearest neighbours for each evaluation.} We used the R-package \textit{Rlof}~\citep{Rlof} for the computation of LOF. 

The second main inspiration for our approach is the \textbf{RobPCA} algorithm by \cite{hubert2005robpca}. The approach employs distances (similar to the proposed core and orthogonal distances) for describing the outlyingness of observations with respect to the majority of the underlying data. This method should work fine with one consistent majority of observations. In the presence of a multigroup structure, we would expect it to fail since the majority of data cannot be properly described with a model of a single normal distribution. RobPCA calculates two outlyingness scores, namely orthogonal and score distances\footnote{For a multivariate interpretation of outlyingness based on those two scores, we refer to \cite{pomerantsev2008acceptance}.}. RobPCA usually flags observations as outliers if either the score distance or the orthogonal distance exceed a certain threshold. This threshold is based on transformations of quantiles of normal and $\chi^2$ distributions. We use the maximum quantile of each observation for the distributions of orthogonal and score distances as a measure for outlyingness in order to stay consistent with the original outlier detection concept of RobPCA. The dimension of the subspace used for dimension reduction is dynamically adjusted. We used the R-package \textit{rrcov}~\citep{rrcov} for the computation of RobPCA.

In addition to LOF and RobPCA, we compare the proposed local projections with \textbf{PCOut} by \cite{filzmoser2008outlier}. PCOut is an outlier detection algorithm where location and scatter outliers are identified based on robust kurtosis and biweight measures of robustly estimated principal components. The dimension of the projection space is automatically selected based on the proportion of the explained overall variance. A combined outlyingness weight is calculated during the process, which we use as an outlyingness score. The method is implemented using the R-package \textit{mvoutlier} \citep{mvoutlier}.

{Another} method included in our comparison is the \textbf{subspace-based outlier detection (SOD)} by \cite{kriegel2009outlier}. The method is looking for relevant dimensions parallel to the axis in which outliers can be identified. The identification of those subspaces is based on knn, where $k$ is optimized in a way similar to  LOF and local projections. 
We used the implementation of SOD in the ELKI framework~\citep{achtert2008elki} for performance reasons.

All three methods, LocOut, LOF and SOD, implement knn-estimations in their respective procedures. Therefore, it is reasonable to monitor the performance of \textbf{k-nearest neighbors (KNN)}, which can be directly used for outlier detection as suggested in \cite{ramaswamy2000efficient}. The performance is optimized over all reasonable $k$ between 1 and the minimal number of non-outlying observations of a group which we take from the ground truth in our evaluation. We used the R-package \textit{dbscan} for the computation of KNN.

Similar to the proposed local projections, \textbf{Outlier Detection in Arbitrary Subspaces (COP)} by~\cite{kriegel2012outlier} locally evaluates the outlyingness of observations. The k-nearest neighbours of each observation are used to estimate the local covariance structure and robustly compute the representation of the evaluated observation in the principal component space. The last principal components are then used to measure the outlyingness, while the number of principal components to be cut off is dynamically computed. {Although} the initial concept looks similar to our proposed algorithm, it contains some disadvantages. The number of observations used for the knn estimation needs to be {a lot} larger than the number of variables. A proportion {of observations to variables} of three to one is suggested. Therefore, the method can not be employed for flat data structures, which represent the focus of the proposed approach for outlier analysis. While COP performed competitive for simulations with no or a very small number of noise variables, the computation of COP is not possible in flat data settings. As the non-flat settings only represent a minor fraction of the overall simulations, we did not include COP in the simulated evaluation but only in the low-dimensional real data evaluation of Section~\ref{sec:application}.

\section{Simulation results}
\label{sec:simulations}

We used two simulation setups to evaluate the performance of the methods for increasing number of noise variables in order to determine their usability for high-dimensional data. We do that by starting with $50$ informative variables and $0$ noise variables, increasing the number of noise variables up to $5000$. We use three groups of observations with 150, 150, and 100 observations. Starting from $350$ noise variables, the data structure becomes flat, which we expect to lead to performance drops as the estimation of the underlying density becomes more and more problematic. Each of the three groups of observations is simulated based on a randomly rotated covariance matrix $\bs \Sigma^i$ as  performed in \cite{campello2015hierarchical},
\begin{align}
\bs \Sigma^i = \begin{pmatrix} 
\bs \Sigma^i_{inf} & \bs 0 \\
\bs 0 & \bs I_{noise} \\
\end{pmatrix} 
&&
\bs \Sigma^i_{inf}  & = \bs \Omega_i \begin{pmatrix}
1 & \rho^i & \dots & \rho^i \\
\rho^i & \ddots  & \ddots & \vdots \\
\vdots & \ddots &  \ddots & \rho^i \\
\rho^i & \dots &  \rho^i & 1\\
\end{pmatrix} \bs \Omega_i ',
\end{align}
for $i=1,2,3$, where $\bs I_{noise}$ is an identity matrix describing the covariance of uncorrelated noise variables and $\bs \Sigma^i_{inf}$ the covariance matrix of informative variables, which are variables containing information about the separation of present groups where $\rho^i$ is randomly selected between 0.1 and 0.9{, $\rho^i \sim U[0.1, 0.9]$}. $\bs \Omega^i$ represents the randomly generated orthonormal rotation matrix. For our simulation setups we always consider the dimensionality of $\bs \Sigma^i_{inf}$ to be $50$. During the simulation, we evaluate the impact of such noise variables and therefore perform the simulation for a varying number of noise variables. While the mean values of the noise variables are fixed to zero for all groups, the mean values of the informative variables are set as follows:
\begin{equation}
(\bs \mu_1, \bs \mu_2, \bs \mu_3) =  \begin{pmatrix}
\mu & 0 & 0 \\
0 & \mu & 0\\
0 & 0 &  \mu \\
 \mu & 0 & 0 \\
0 & \mu & 0\\
\vdots & \vdots& \vdots\\
\end{pmatrix}.
\end{equation}
Therefore, for each informative variable, one group can be distinguished from the two other groups. The degree of separation, given by $\mu$, is randomly selected from a uniform distribution $U_{[ -6,-3 ] \cup [ 3,6 ]}$. The first simulation setup uses multivariate normally distributed groups of observations using the parameters $\bs \mu_i$ and $\bs \Sigma_{inf}^i$, for $i \in \{ 1,2,3 \}$, and the second setup uses multivariate log-normally distributed groups of observations with the same parameters. {Note that noise variables can be problematic for several of the outlier detection methods, and skewed distributions can create difficulties for methods relying on elliptical distributions.} 

After simulating the groups of observations, scatter outliers are generated by replacing 5\% of the observations of each group with outlying observations. Therefore, we use the same location parameter $\bs \mu_i$, but their covariance matrix is a diagonal matrix with constant diagonal elements {$\sigma$} which are randomly generated {between $3$ and $9$, $\sigma \sim U[3,9],$} for informative variables. The reason for using scatter outliers instead of location outliers (changed $\bs \mu_i$) is the advantage, that outliers will not form a separate group but will stick out of their respective group in random directions.

\begin{figure}[!htb]
\subfloat[]{  
    \includegraphics[width=0.38\linewidth, valign=t]{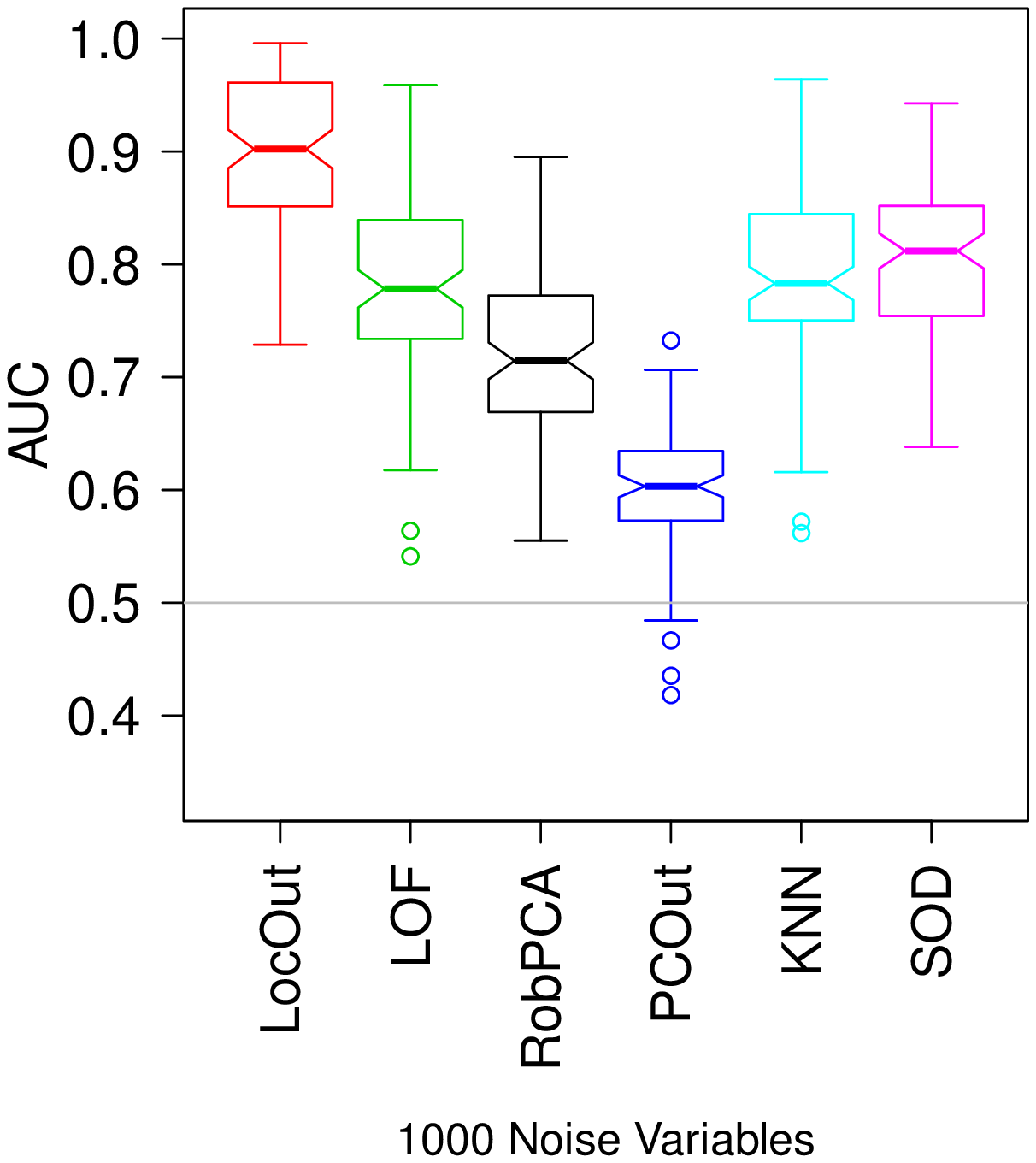}
    \label{fig:sim1a}}
\subfloat[]{  
    \includegraphics[width=0.6\linewidth, valign=t]{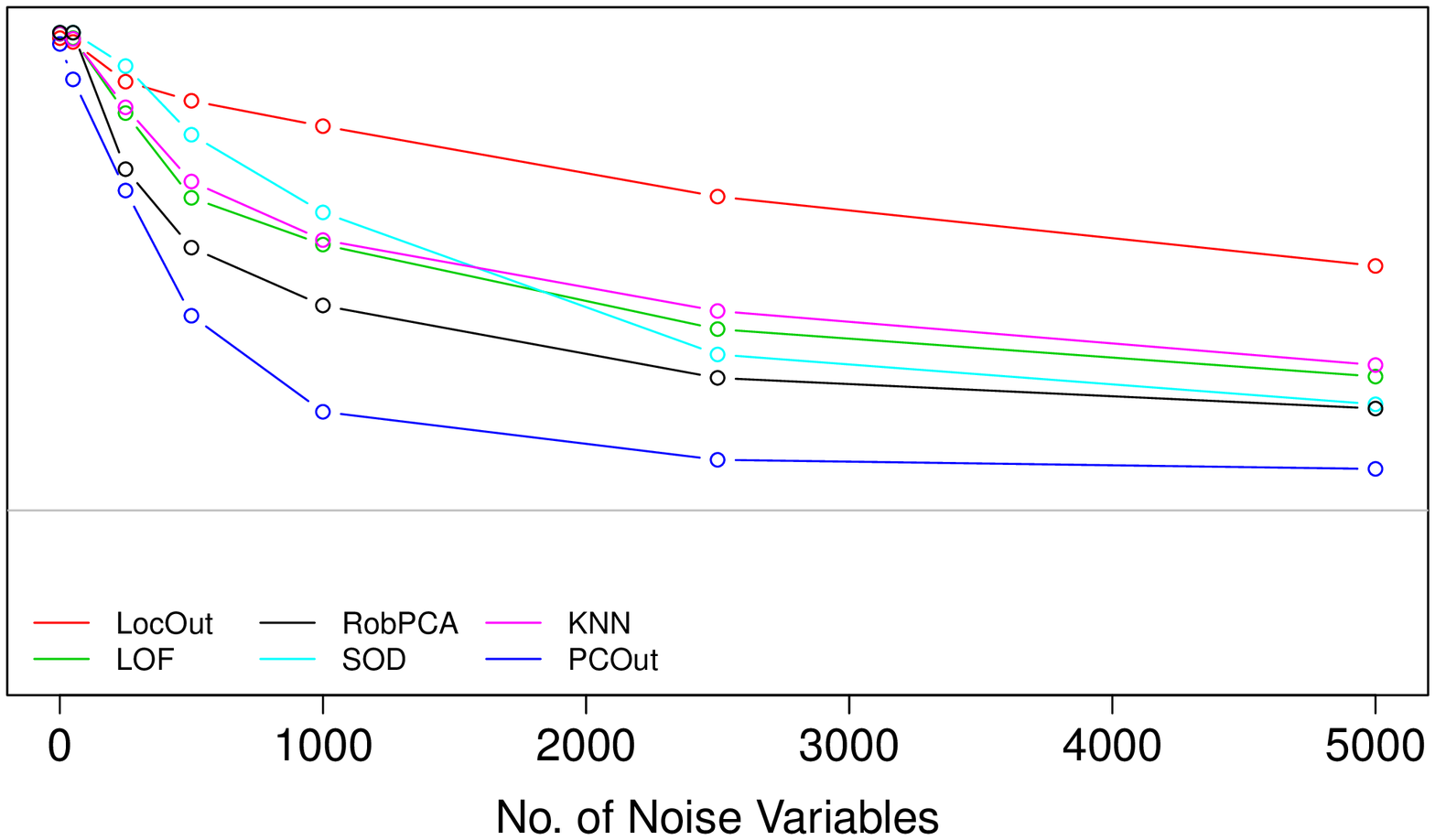} 
    \label{fig:sim1b}}
\caption{Evaluation of outliers in three multivariate normally distributed groups with a varying number of noise variables. 5\% of the observations were replaced by outliers. Plot (a) shows boxplots for the setup with 1000 noise variables. Each setup was repeatedly simulated 100 times. Plot (b) shows the median performance of each method for various numbers of noise variables.}
\label{fig:sim1}
\end{figure}

The outcome of the first simulation setup based on multivariate normal distribution is visualized in Figure~\ref{fig:sim1}. {Figure~\ref{fig:sim1a} shows the performance for 100 repetitions with 1000 noise variables as boxplots measured by the AUC value. We note that local projections (LocOut)  outperform all other methods, while LOF, SOD, and KNN perform approximately at the same level. For smaller numbers of noise variables, especially SOD performs better than local projections.} This becomes clear in Figure~\ref{fig:sim1b}, showing the median performance of all methods with a varying number of noise variables. {We see that the performance of SOD drops quicker than other methods, while local projections are effected the least by an increasing number of noise variables.}
The horizontal grey line {corresponds} to a performance of 0.5 which refers to random outlier scores.

\begin{figure}[!hbt]
\centering
\subfloat[]{  
    \includegraphics[width=0.38\linewidth, valign=t]{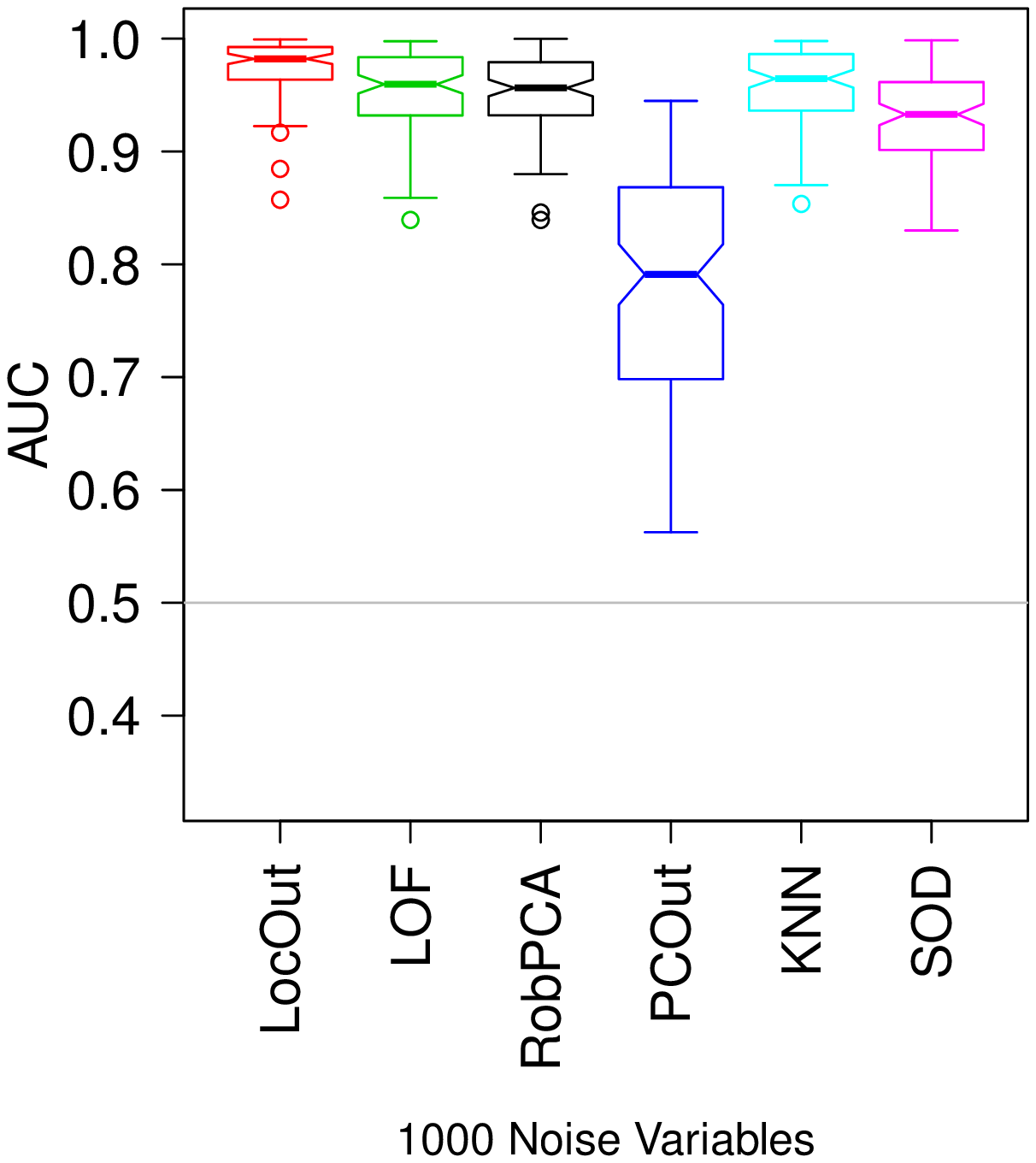} \label{fig:sim2a}}
\subfloat[]{  
    \includegraphics[width=0.6\linewidth, valign=t]{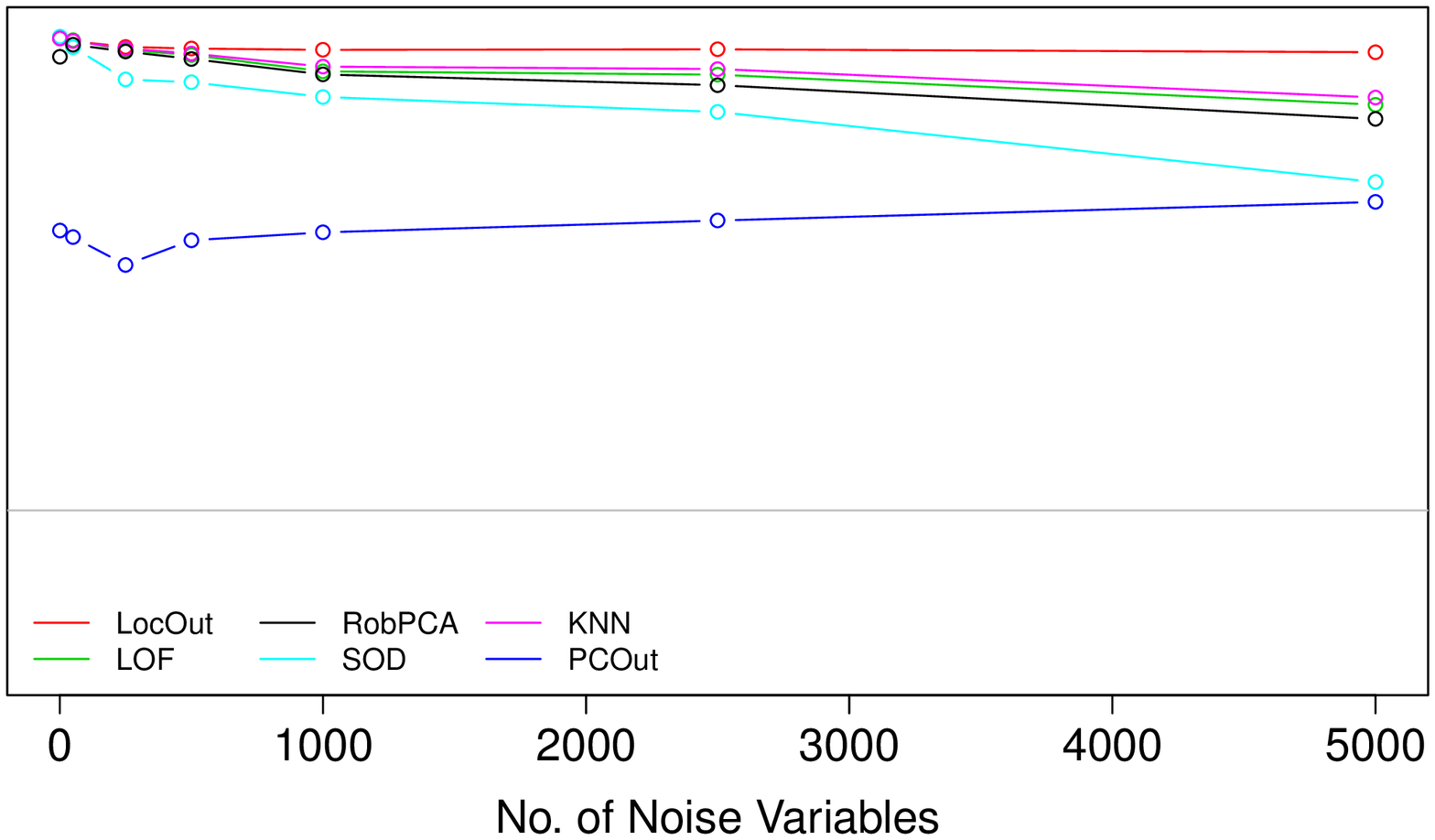} 
    \label{fig:sim2b}}
\caption{Evaluation of outliers in three multivariate log-normally distributed groups with a varying number of noise variables. 5\% of the observations were replaced by outliers. Plot (a) shows boxplots for the setup with 1000 noise variables. Each setup was repeatedly simulated 100 times. Plot (b) shows the median performance of each method for various numbers of noise variables.}
\label{fig:sim2}
\end{figure}

Setup 2, visualized in Figure~\ref{fig:sim2}, shows the effect of non-normal distributions on the outlier detection methods. The same  parameters used for log-normal distributions as in the normally distributed setup, make it easier {for all methods} to identify outliers. {Nevertheless, the order of performance changes since the methods are affected differently.}  SOD is stronger affected than LOF, since it is easier for SOD to identify useful spaces for symmetric distributions while LOF does not benefit from such properties. LocOut still shows the best performance, at least for an increasing number of noise variables. {The most notable difference is the effect on RobPCA, which heavily depends on the violated assumption of normal distribution.}

\section{Application on real-world datasets}
\label{sec:application}

{In order to demonstrate the effectiveness of local projections in real-world applications, we analyze three different datasets, varying in the number of groups, the dimension of the data space, and  the separability of the groups. We always use observations from multiple groups as non outlying observations and a small number of one additional group to simulate outliers in the dataset.}

\subsection{Olive Oil}

The first real-world dataset consists of 120 samples of measurements of 25 chemical compositions (fatty acids, sterols, triterpenic alcohols) of olive oils from Tuscany, Italy \citep{armanino1989chemometric}. The dataset is used as a reference dataset in the R-package \textit{rrcovHD} \citep{rrcovHD} for robust multivariate methods for high-dimensional data and consists of four groups of 50, 25, 34, and 11 observations, respectively. We use observations from the smallest group with 11 observations to sample $5$ outliers 50 times.

In our context, this dataset represents a situation where the distribution can be well-estimated due to its non-flat data structure. Therefore, it is possible to include COP in the evaluation. It is important to note that at least 26 observations must be used by COP in order to be able to locally estimate the covariance structure, while there will always be a smallest group of 25 observations at most present for each setup. Thus, we would assume, that COP has problems distinguishing between outliers and observations from this smallest group which does not yield enough observations for the covariance estimation.

\begin{figure}[tbh]
\centering
\includegraphics[width=0.475\linewidth]{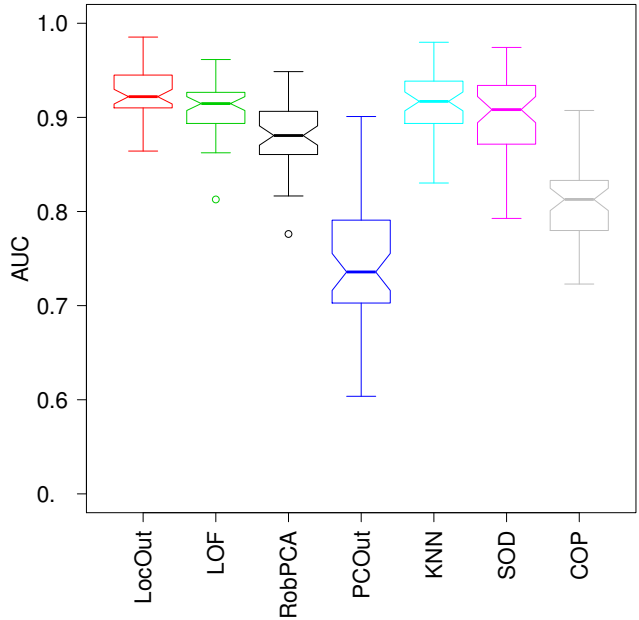}
\caption{Performance of different outlier detection methods for the 25 dimensional olive oil dataset measured by the area under the ROC curve (AUC). For each method the configuration parameters are optimized based on the ground truth.}
\label{fig:evalOlitos}
\end{figure}

{We show the performance of the compared outlier detection methods based on the AUC values in Figure~\ref{fig:evalOlitos}. We note that all methods but PCOut and COP perform at a very high level. For KNN, SOD and LocOut, there is only a non-significant difference in the median performance.}

\subsection{Melon}

{The second real-world dataset used for the evaluation is a fruit data set, which consists of 1095 observations in a 256 dimensional space corresponding to spectra of the different melon cultivars. The observations are documented as members of three groups of sizes 490, 106 and 499,} but in addition, during the cultivation processes different illumination systems have been used leading to subgroups. The dataset is often used to evaluate robust statistical methods \citep[e.g.][]{hubert2004fast}.

\begin{figure}[!htb]
\centering
\includegraphics[width=0.475\linewidth]{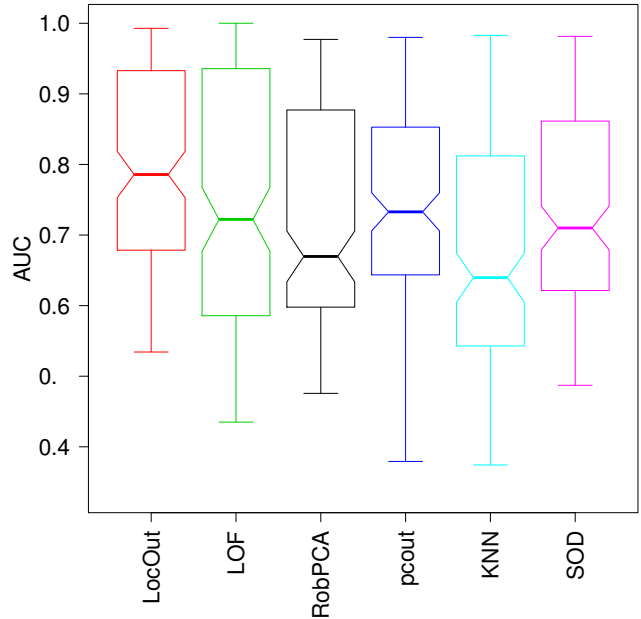} 
\caption{Evaluation of the performance of the outlier detection algorithms on the fruit data set,  showing boxplots of the performance of 150 repetitions of outlier detection measured by the AUC.}
\label{fig:fruit}
\end{figure}

{We sample 100 observations from two randomly selected main groups to simulate a highly inconsistent structure of main observations and add 7 outliers, randomly selected from the third remaining group. We repeatedly simulate such a setup 150 times in order to make up for the high degree of inconsistency. As Figure~\ref{fig:fruit} shows, the identification of outliers is extremely difficult for this dataset. A combination of properly reducing the dimensionality and modelling the existing sub-groups is required. LocOut outperforms the compared methods, followed by LOF and PCOut.}

\subsection{Archaeological glass vessels}

{The observations of the glass vessels dataset, described e.g.~in \cite{janssens1998composition} refer to} archaeological glass vessels, which have been excavated in Antwerp (Belgium).  In order to distinguish between different types of glass, which have either been imported or locally produced, 180 glass samples were chemically analyzed using electron probe X-ray microanalysis (EPXMA). By measuring a spectrum at 1920 different energy levels corresponding to different chemical elements for each of those vessels, a high-dimensional data set for classifying the glass vessels is created. A few {(11)} of those variables/energy levels contain no variation and are therefore removed from our experiments in order to avoid problems during the computation of outlyingness. 

\begin{figure}[!htb]
\centering
\subfloat[]{ 
    \includegraphics[width=0.457\linewidth, valign=t]{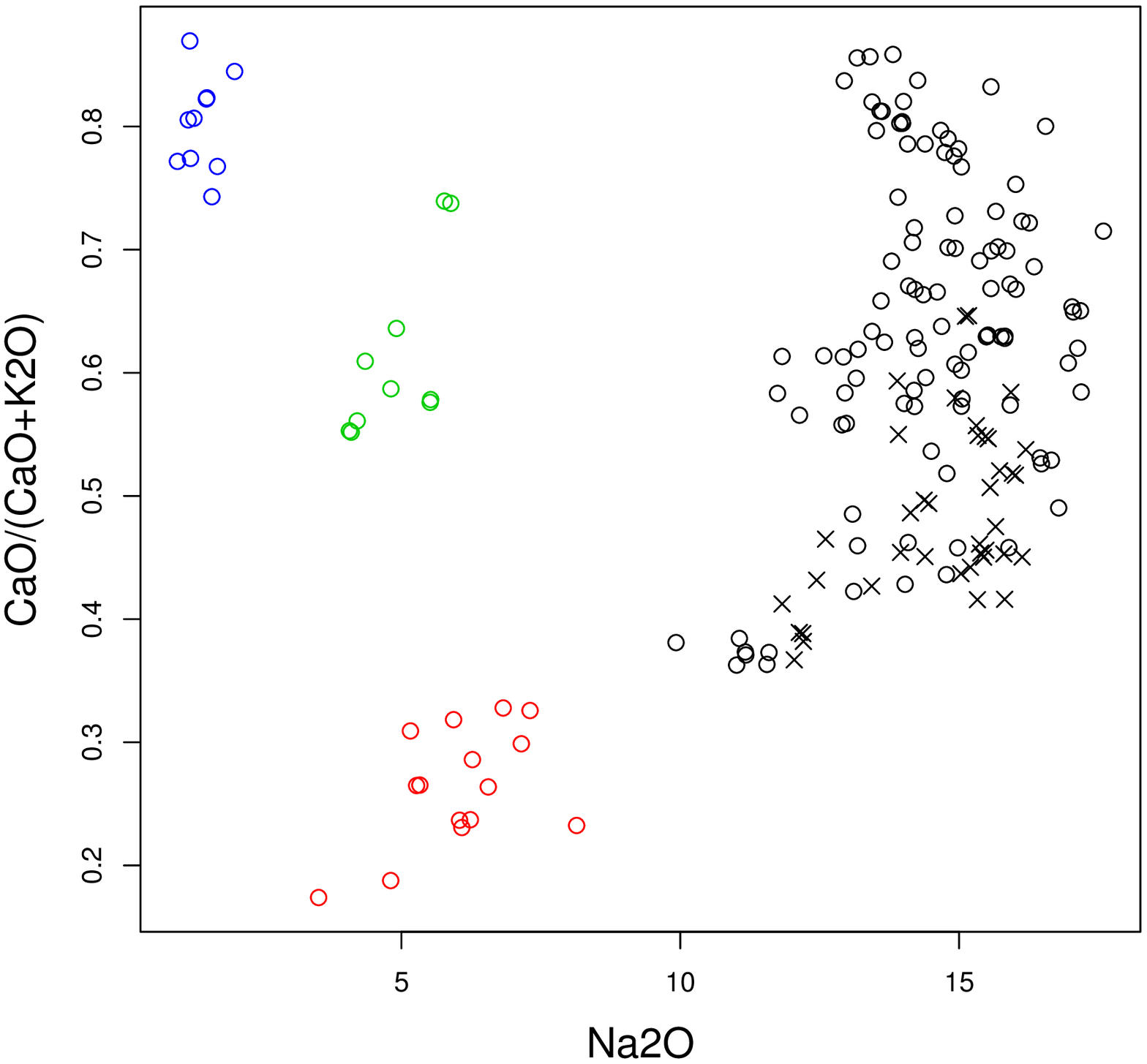} 
    \label{fig:glass1a}}
\subfloat[]{ 
    \includegraphics[width=0.457\linewidth , valign=t]{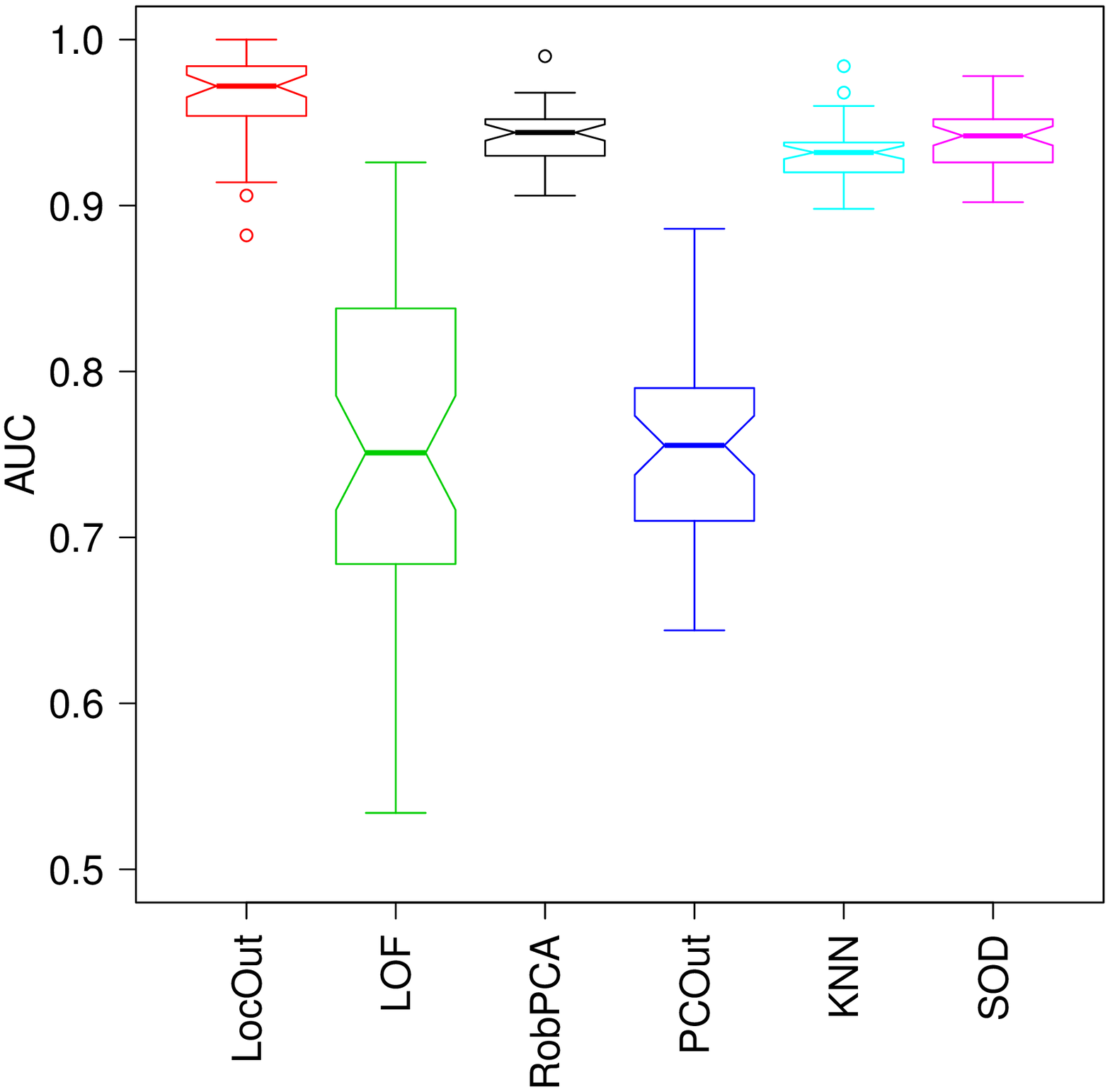} 
    \label{fig:glass1b}}
\caption{Evaluation of the performance of the outlier detection algorithms on the glass vessels data set. Plot (a) shows the classification of the group structure based on the chemical composition. Plot (b) shows boxplots of the performance of 50 repetitions of outlier detection measured by the AUC.}
\label{fig:glass1}
\end{figure}

While performing PLS regression analysis, \cite{lemberge2000quantitative} realized that some vessels had been measured at a different detector efficiency and, therefore, removed those spectra from the dataset. We do not remove those observations, since from an outlier detection perspective they represent bad leverage points as indicated by \cite{serneels2005partial}, which we want to be able to identify. These leverage points are visualized in Figure~\ref{fig:glass1a} with x-symbols. By including these observations as part of the main groups, it becomes especially difficult to identify outliers sampled from the green group (potasso-calic). We sample 100 observations from the non-potasso-calic group 50 times and add 5 randomly selected potasso-calic observations as outliers. The performance is visualized in Figure~\ref{fig:glass1b}. Again, LocOut outperforms all compared methods, while LOF and PCOut have problems to deal with this data setup.

\section{Discussion of runtime}
\label{sec:computational}

The algorithm for local projections was implemented in an R-package which is publicly available\footnote{http://www.applied-statistics.at/locout\_1.0.tar.gz}. The package further includes the glass vessels data set used in Section~\ref{sec:application}. Based on this R-package, we performed simulations to test the required computational effort for the proposed algorithm and the impact of changes in the number of observations and the number of variables.

For each projection, the first step of our proposed algorithm is based on the $k$-nearest neighbours  concept. Therefore, we need to compute the distance matrix for all $n$ available $p$-dimensional observations leading to an effort of $O(n(n-1)p/2)$, where $n(n-1)/2$ represents the combinations of observations and $p$, being the dimension of the data space, reflects the effort for the computation of each Euclidean distance. 

After the basic distance computation, we need to compare those distances (which scales with $n$ but only contributes negligibly to the overall effort) and scale the data based on the location and scale estimation for the selected core (which also does not significantly affect the computation time). 

\begin{figure}[!h]
\centering
\subfloat[]{  \includegraphics[width=0.457\linewidth]{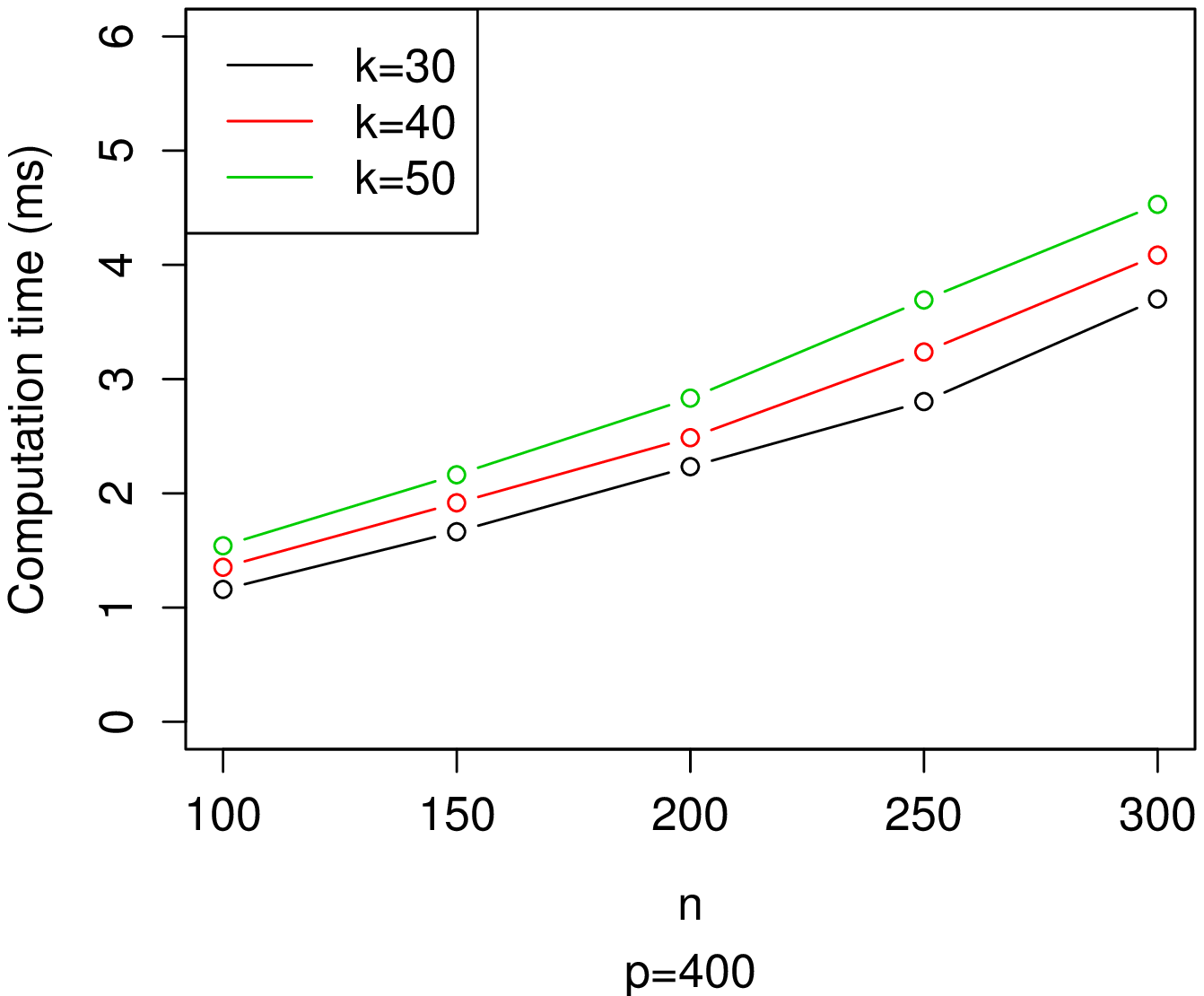} }
\subfloat[]{         \includegraphics[width=0.457\linewidth]{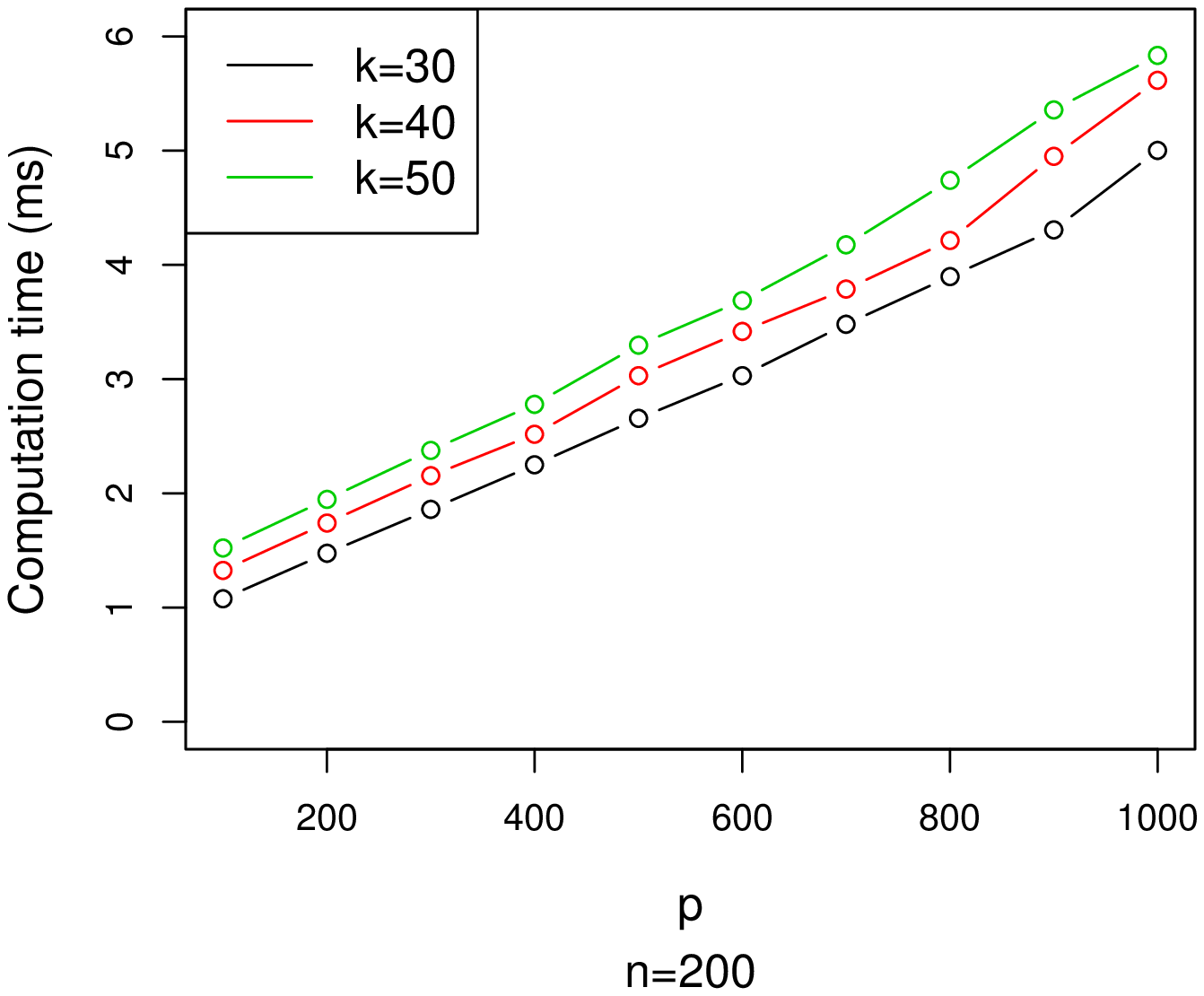} } \\
\subfloat[]{         \includegraphics[width=0.457\linewidth]{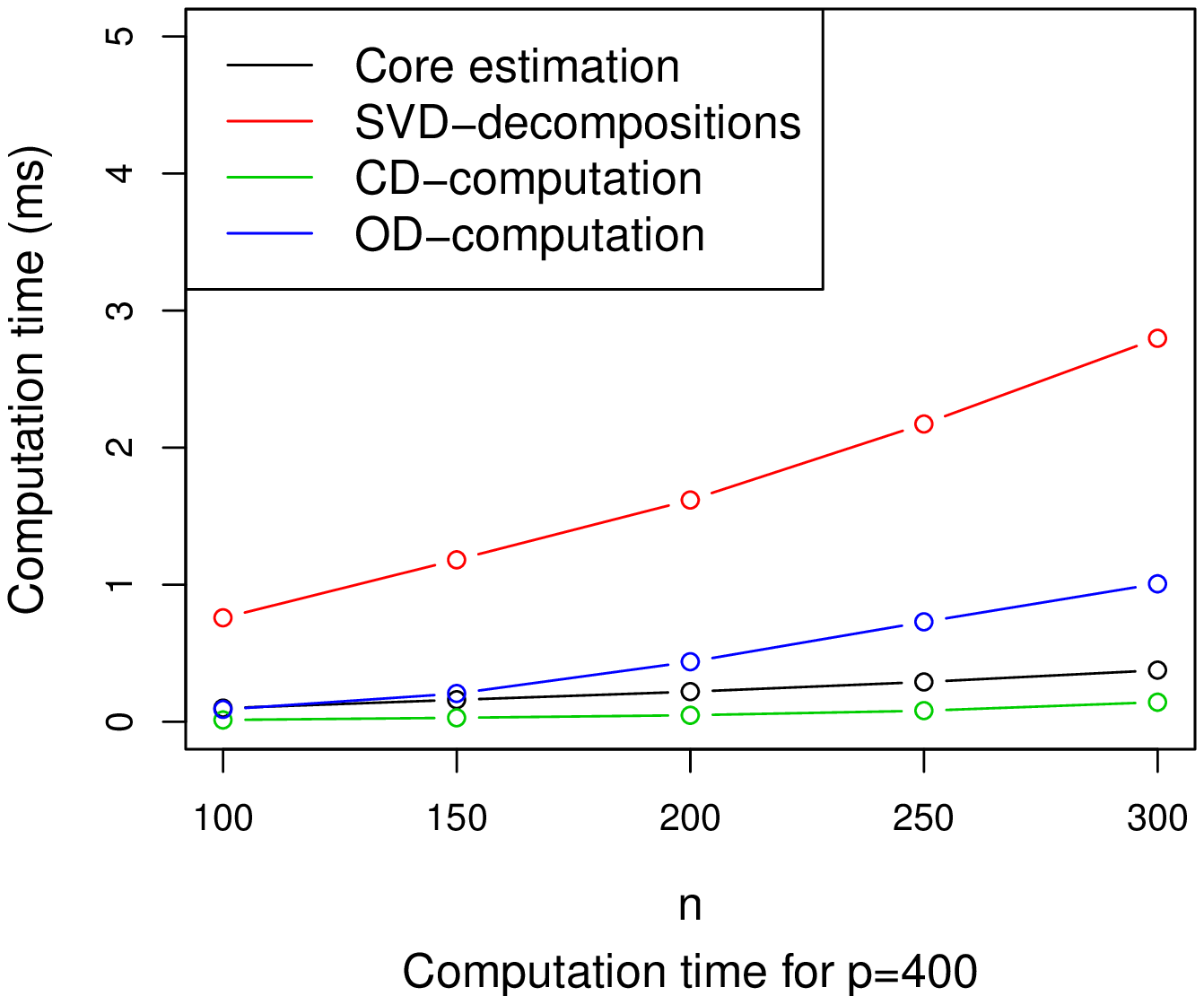} }
\subfloat[]{         \includegraphics[width=0.457\linewidth]{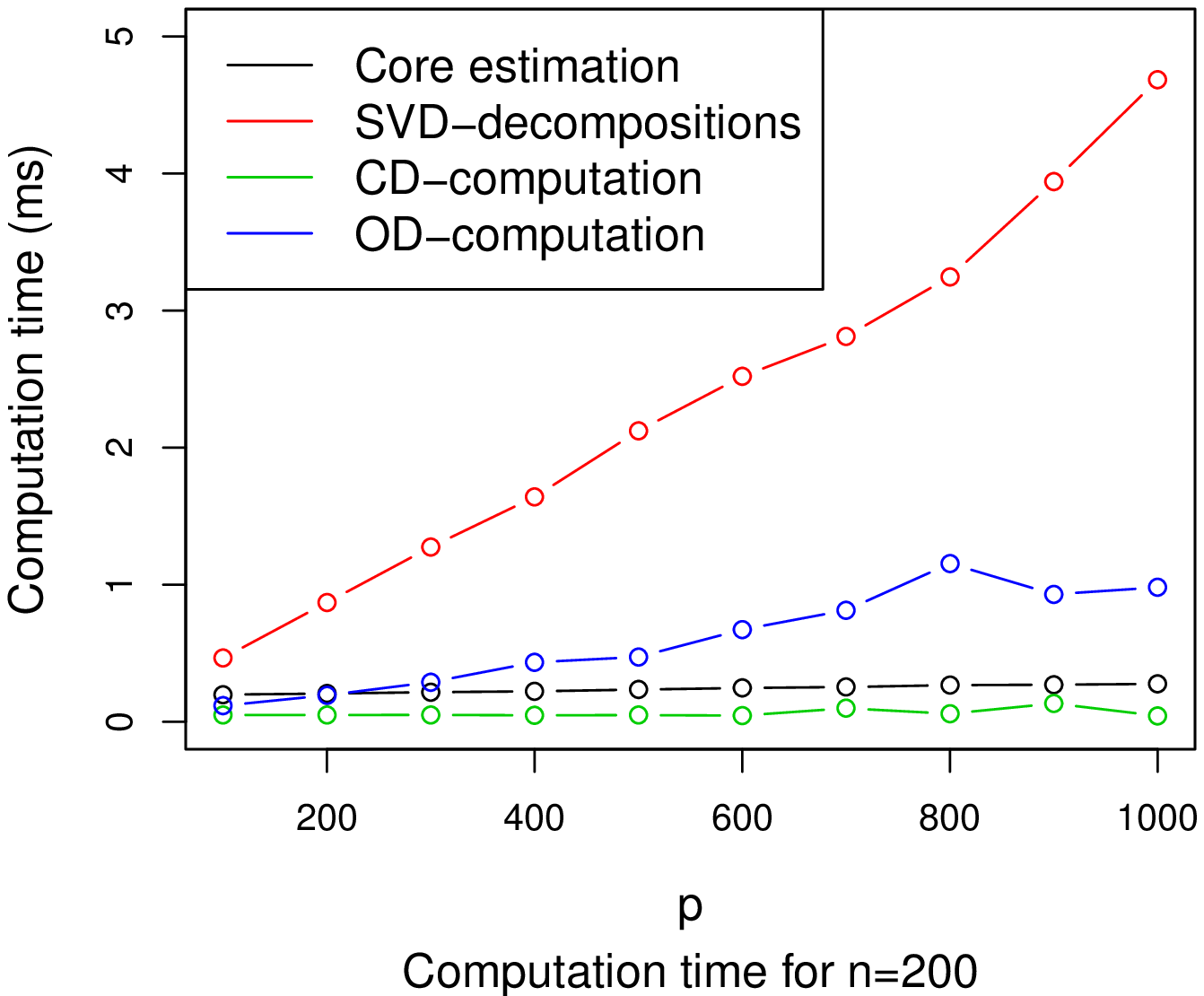} }
\caption{Visualisation of the computation time of the local projections. Plots (a) and (b) evaluate the development of the overall computation time for increasing $n$ in plot (a) and increasing $p$ in plot  (b). Those evaluations are performed for varying $k$.  Plot(c) and (d) focus on different components of the computation for a fixed $k=40$ and increasing $n$ and $p$.}
\label{fig:computationTime}
\end{figure}

For all of the $n$ cores, we perform an SVD decomposition leading to an effort of $O(p^2n^2+n^4)$. Therefore, a total effort of $O(n^2p(1+p) + n^4)$ is expected for the computation of all local projections. In this {calculation}, reductions, such as the multiple computation of the projection onto the same core, are not taken into account. Such an effect is very common due to the presence of hubs in data sets \citep{zimek2012survey}. Figure~\ref{fig:computationTime} provide an overview of the overall computation time decomposed into the different aspects of the computation algorithm.

We observe that the computation time increases approximately linearly with $p$, while it increases faster than a linear term with increasing $n$. There is an interaction effect between $k$ and $n$ visible in plot (a) of Figure~\ref{fig:computationTime} as well, due to the necessity of $n$ knn computations. Plots (c) and (d) show that the {key factors} are the $n$ SVDs. Especially the core estimation and the computation of the core distance are just marginally affected by increasing $n$ and not affected at all by increasing $p$. The orthogonal distance computation is non-linearly affected by increasing $n$ and $p$ which however remains relatively small when being compared to the SVD estimations.

\section{Conclusions}
\label{sec:conclusion}

We proposed a novel approach for evaluating the outlyingness of observations based on their local behaviour, named \textit{local projections}. By combining techniques from the existing robust outlier detection RobPCA \citep{hubert2005robpca} and from Local Outlier Factor (LOF) \citep{breunig2000lof}, we created a method for outlier detection, which is highly robust towards large numbers of non-informative noise variables and which is able to deal with multiple groups of observations, not necessarily following any specific standard distribution.

These properties are gained by creating a local description of a data structure by robustly selecting a number of observations based on the k-nearest neighbours of an initiating observation and projecting all observations onto the space spanned by those observations. Doing so repeatedly, where each available observation initiates a local description, we describe the full space in which the data set is located. In contrast to existing subspace-based methods, we create a new concept for interpreting the outlyingness of observations with respect to such a projection by introducing the concept of quality of local description of a model for outlier detection. By aggregating the measured outlyingness of each projection and by downweighting the outlyingness with this quality-measure of local description, we define the univariate local outlyingness score, \textit{LocOut}. \textit{LocOut} measures the outlyingness of each observation in comparison to other observations and results in a ranking of outlyingness for all observations. We do not provide cut off values for classifying observations as outliers and non-outliers. While at first consideration this poses a disadvantage, it allows for disregarding any assumptions about the data distribution. Such assumptions would be required in order to compute theoretical critical values. 

We showed that this approach is more robust towards the presence of non-informative noise variables in the data set than other well-established methods we compared to (LOF, SOD, PCOut, KNN, COP, and RobPCA). Additionally, skewed non-symmetric data structures have less impact than for the compared methods. These properties, in combination with the new interpretation of outlyingness allowed for a competitive analysis of high-dimensional data sets as demonstrated on {three real-world application of varying dimensionality and group structure.}

The overall concept of the proposed local projections utilized for outlier detection opens up possibilities for more general data analysis concepts. Any clustering method and discriminant analysis method is based on the idea of observations being an outlier for one group and therefore being part of another group.  By combining the different local projections, a possibility for avoiding assumptions about the data distribution - which are in reality often violated - is provided. Thus, applying local projections on data analysis problems could not only provide a suitable method for analyzing high-dimensional problems but could also reveal additional information on method-influencing observations due to the quality of local description interpretation of local projections.

\section*{Acknowledgements}
This work has been partly funded by the Vienna Science and Technology Fund (WWTF) through project ICT12-010 and by the K-project DEXHELPP through COMET - Competence Centers for Excellent Technologies, supported by BMVIT, BMWFW and the province Vienna. The COMET program is administrated by FFG.

\bibliographystyle{apalike}

\begin{thebibliography}{}

\bibitem[Achtert et~al., 2008]{achtert2008elki}
Achtert, E., Kriegel, H.-P., and Zimek, A. (2008).
\newblock Elki: a software system for evaluation of subspace clustering
  algorithms.
\newblock In {\em Scientific and statistical database management}, pages
  580--585. Springer.

\bibitem[Aggarwal and Yu, 2001]{aggarwal2001outlier}
Aggarwal, C.~C. and Yu, P.~S. (2001).
\newblock Outlier detection for high dimensional data.
\newblock In {\em ACM Sigmod Record}, volume~30, pages 37--46. ACM.

\bibitem[Armanino et~al., 1989]{armanino1989chemometric}
Armanino, C., Leardi, R., Lanteri, S., and Modi, G. (1989).
\newblock Chemometric analysis of tuscan olive oils.
\newblock {\em Chemometrics and Intelligent Laboratory Systems}, 5(4):343--354.

\bibitem[Breunig et~al., 2000]{breunig2000lof}
Breunig, M.~M., Kriegel, H.-P., Ng, R.~T., and Sander, J. (2000).
\newblock Lof: identifying density-based local outliers.
\newblock In {\em ACM sigmod record}, volume~29, pages 93--104. ACM.

\bibitem[Campello et~al., 2015]{campello2015hierarchical}
Campello, R.~J., Moulavi, D., Zimek, A., and Sander, J. (2015).
\newblock Hierarchical density estimates for data clustering, visualization,
  and outlier detection.
\newblock {\em ACM Transactions on Knowledge Discovery from Data (TKDD)},
  10(1):5.

\bibitem[Campos et~al., 2016]{campos2016evaluation}
Campos, G.~O., Zimek, A., Sander, J., Campello, R.~J., Micenkov{\'a}, B.,
  Schubert, E., Assent, I., and Houle, M.~E. (2016).
\newblock On the evaluation of unsupervised outlier detection: measures,
  datasets, and an empirical study.
\newblock {\em Data Mining and Knowledge Discovery}, 30(4):891--927.

\bibitem[De~Maesschalck et~al., 2000]{de2000mahalanobis}
De~Maesschalck, R., Jouan-Rimbaud, D., and Massart, D.~L. (2000).
\newblock The mahalanobis distance.
\newblock {\em Chemometrics and intelligent laboratory systems}, 50(1):1--18.

\bibitem[Fawcett, 2006]{fawcett2006introduction}
Fawcett, T. (2006).
\newblock An introduction to roc analysis.
\newblock {\em Pattern recognition letters}, 27(8):861--874.

\bibitem[Filzmoser and Gschwandtner, 2015]{mvoutlier}
Filzmoser, P. and Gschwandtner, M. (2015).
\newblock {\em mvoutlier: Multivariate outlier detection based on robust
  methods}.
\newblock R package version 2.0.6.

\bibitem[Filzmoser et~al., 2008]{filzmoser2008outlier}
Filzmoser, P., Maronna, R., and Werner, M. (2008).
\newblock Outlier identification in high dimensions.
\newblock {\em Computational Statistics \& Data Analysis}, 52(3):1694--1711.

\bibitem[Henrion et~al., 2013]{henrion2013casos}
Henrion, M., Hand, D.~J., Gandy, A., and Mortlock, D.~J. (2013).
\newblock Casos: a subspace method for anomaly detection in high dimensional
  astronomical databases.
\newblock {\em Statistical Analysis and Data Mining: The ASA Data Science
  Journal}, 6(1):53--72.

\bibitem[Hu et~al., 2015]{Rlof}
Hu, Y., Murray, W., Shan, Y., and {Australia.} (2015).
\newblock {\em Rlof: R Parallel Implementation of Local Outlier Factor(LOF)}.
\newblock R package version 1.1.1.

\bibitem[Hubert et~al., 2005]{hubert2005robpca}
Hubert, M., Rousseeuw, P.~J., and Vanden~Branden, K. (2005).
\newblock Robpca: a new approach to robust principal component analysis.
\newblock {\em Technometrics}, 47(1):64--79.

\bibitem[Hubert and Van~Driessen, 2004]{hubert2004fast}
Hubert, M. and Van~Driessen, K. (2004).
\newblock Fast and robust discriminant analysis.
\newblock {\em Computational Statistics \& Data Analysis}, 45(2):301--320.

\bibitem[Janssens et~al., 1998]{janssens1998composition}
Janssens, K., Deraedt, I., Freddy, A., and Veekman, J. (1998).
\newblock Composition of 15-17th century archeological glass vessels excavated
  in antwerp. belgium.
\newblock {\em Mikrochimica Acta. v15 iSuppl}, pages 253--267.

\bibitem[Kriegel et~al., 2009]{kriegel2009outlier}
Kriegel, H.-P., Kr{\"o}ger, P., Schubert, E., and Zimek, A. (2009).
\newblock Outlier detection in axis-parallel subspaces of high dimensional
  data.
\newblock {\em Advances in knowledge discovery and data mining}, pages
  831--838.

\bibitem[Kriegel et~al., 2012]{kriegel2012outlier}
Kriegel, H.-P., Kroger, P., Schubert, E., and Zimek, A. (2012).
\newblock Outlier detection in arbitrarily oriented subspaces.
\newblock In {\em Data Mining (ICDM), 2012 IEEE 12th International Conference
  on}, pages 379--388. IEEE.

\bibitem[Lemberge et~al., 2000]{lemberge2000quantitative}
Lemberge, P., De~Raedt, I., Janssens, K.~H., Wei, F., and Van~Espen, P.~J.
  (2000).
\newblock Quantitative analysis of 16--17th century archaeological glass
  vessels using pls regression of epxma and $\mu$-xrf data.
\newblock {\em Journal of Chemometrics}, 14(5-6):751--763.

\bibitem[Ortner et~al., 2017]{ortner2017guided}
Ortner, T., Filzmoser, P., Zaharieva, M., Breiteneder, C., and Brodinova, S.
  (2017).
\newblock Guided projections for analysing the structure of high-dimensional
  data.
\newblock {\em arXiv preprint arXiv:1702.06790}.

\bibitem[Pomerantsev, 2008]{pomerantsev2008acceptance}
Pomerantsev, A.~L. (2008).
\newblock Acceptance areas for multivariate classification derived by
  projection methods.
\newblock {\em Journal of Chemometrics}, 22(11-12):601--609.

\bibitem[Ramaswamy et~al., 2000]{ramaswamy2000efficient}
Ramaswamy, S., Rastogi, R., and Shim, K. (2000).
\newblock Efficient algorithms for mining outliers from large data sets.
\newblock In {\em ACM Sigmod Record}, volume~29, pages 427--438. ACM.

\bibitem[Serneels et~al., 2005]{serneels2005partial}
Serneels, S., Croux, C., Filzmoser, P., and Van~Espen, P.~J. (2005).
\newblock Partial robust m-regression.
\newblock {\em Chemometrics and Intelligent Laboratory Systems}, 79(1):55--64.

\bibitem[Todorov, 2016]{rrcovHD}
Todorov, V. (2016).
\newblock {\em rrcovHD: Robust Multivariate Methods for High Dimensional Data}.
\newblock R package version 0.2-5.

\bibitem[Todorov and Filzmoser, 2009]{rrcov}
Todorov, V. and Filzmoser, P. (2009).
\newblock An object-oriented framework for robust multivariate analysis.
\newblock {\em Journal of Statistical Software}, 32(3):1--47.

\bibitem[Zimek et~al., 2012]{zimek2012survey}
Zimek, A., Schubert, E., and Kriegel, H.-P. (2012).
\newblock A survey on unsupervised outlier detection in high-dimensional
  numerical data.
\newblock {\em Statistical Analysis and Data Mining: The ASA Data Science
  Journal}, 5(5):363--387.

\end{thebibliography}

\end{document}